\documentclass[prx, preprint, superscriptaddress]{revtex4-2}
\usepackage{graphicx}
\usepackage{amssymb}
\usepackage{amsfonts}
\usepackage{amsmath}
\usepackage{lmodern}
\usepackage{mathrsfs}
\usepackage{mathtools, slashed}
\usepackage{mathdots}
\usepackage{collref}
\usepackage[colorlinks=true, citecolor=blue, urlcolor=blue]{hyperref}
\usepackage{tikz}
\usepackage{bm}
\usepackage[markup=underlined]{changes}
\makeatletter
\AddToHook{cmd/added/before}{\def\Changes@AuthorColor{purple}}
\AddToHook{cmd/deleted/before}{\def\Changes@AuthorColor{green}}
\AddToHook{cmd/replaced/before}{\def\Changes@AuthorColor{red}}
\makeatother
\usepackage{todonotes}
\setcommentmarkup{\todo[color={pink}, size=\scriptsize]{#3: #1}}

\usetikzlibrary{shapes.geometric}

\usepackage{multirow}

\usepackage{cleveref}
\usepackage[all]{hypcap}
\allowdisplaybreaks

\begin{document}
\title{Squeezed superradiant lasing of a quantum many-body {emitter}}
\author{Da-Wu Xiao}\author{Chong Chen}
\affiliation{Department of Physics, The Chinese University of Hong Kong, Shatin, New Territories, Hong Kong, China}%
\affiliation{The State Key Laboratory of  Quantum Information Technologies and Materials, The Chinese University of Hong Kong, Shatin, New Territories, Hong Kong, China}%
\affiliation{New Cornerstone Science Laboratory, The Chinese University of Hong Kong, Shatin, New Territories, Hong Kong, China}%

\author{Ren-Bao Liu}
\email{rbliu@cuhk.edu.hk}
\affiliation{Department of Physics, The Chinese University of Hong Kong, Shatin, New Territories, Hong Kong, China}%
\affiliation{The State Key Laboratory of  Quantum Information Technologies and Materials, The Chinese University of Hong Kong, Shatin, New Territories, Hong Kong, China}%
\affiliation{New Cornerstone Science Laboratory, The Chinese University of Hong Kong, Shatin, New Territories, Hong Kong, China}%
\affiliation{The Hong Kong Institute of Quantum Information Science and Technology, The Chinese University of Hong Kong, Shatin, New Territories, Hong Kong, China}%
\affiliation{Centre for Quantum Coherence, The Chinese University of Hong Kong, Shatin, New Territories, Hong Kong, China}%

\date{\today}

\begin{abstract}
    In conventional lasers, the emitters are typically incoherent, radiating photons independently; in superradiant lasers, many coherent emitters radiate photons collectively, but they essentially do not interact with each other. Here, we present the concept of quantum many-body lasers, in which the emitters interact coherently and radiate collectively. In this proof-of-concept study, we consider a cavity coupled to many pumped spin-1/2 emitters with all-to-all interaction. We find that the squeezing induced by the coherent many-body interaction can be transferred from the spins to photons through  superradiant lasing. This work illustrates the concept of using a pumped quantum many-body system to generate bright quantum light with quantum correlations beyond conventional optical coherence, which can facilitate quantum technologies and the study of nonlinear optics in the quantum realm.
\end{abstract}
\maketitle

\begin{figure}[t]
\includegraphics[width=0.9\textwidth]{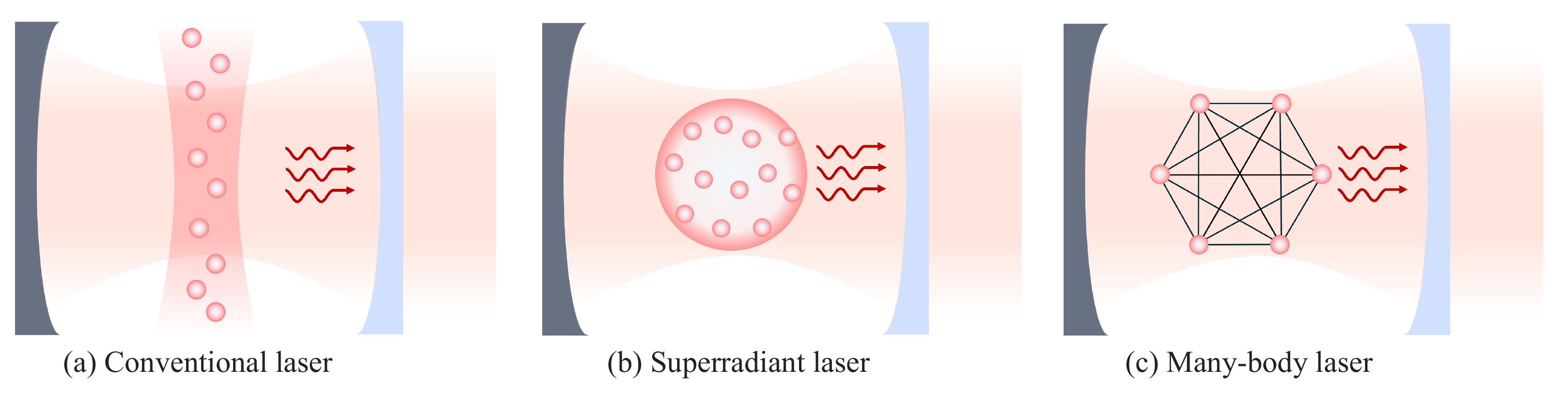}
\caption{\label{fig:figure1}
Schematics of different types of lasers. 
(a) In a conventional laser, population-inverted emitters are independently injected into the cavity. 
(b) In a superradiant laser, many non-interacting emitters couple collectively to a cavity. 
(c) In a quantum many-body laser (proposed in this work), many spins interact coherently and couple collectively to a cavity and the photons are emitted by transitions between quantum many-body states (which can have non-trivial correlations). 
}
\end{figure}

Photons virtually have no mutual interaction, and therefore the coherence and correlations among them, once generated, can be long-lived. 
Such coherence and correlations are resources for metrology~\cite{macek1963rotation}, communication~\cite{kao1966dielectric}, optical computing~\cite{sawchuk1984digital}, and even quantum computing~\cite{knill2001scheme, zhong2020quantum}.
Lasing is the only method to produce a macroscopic number of photons with good coherence. 
In conventional lasers, randomly injected emitters have no coherence or correlation (Fig.~\ref{fig:figure1}a), and it is the population of many photons in a good cavity that enhances the optical coherence~\cite{schawlow1958infrared, maiman1960laser, scully1967quantum}.
When the emitters have good coherence, they can radiate photons collectively, leading to superradiant lasers (Fig.~\ref{fig:figure1}b), which can have good optical coherence even in the bad cavity limit~\cite{dicke1964coherence, haake1993superradiant, meiser2009prospects, bohnet2012steady}.
Typically, the emitters in superradiant lasers do not interact coherently with each other, although incoherent effects such as inhomogeneous broadening and decoherence have been considered~\cite{xu2014synchronization}.

Quantum correlations beyond the optical coherence, such as entanglement and squeezing, are crucial for applications in quantum metrology~\cite{caves1981quantum, yurke19862, dowling2008quantum}, quantum communication~\cite{liao2017satellite}, optical quantum computing~\cite{zhong2020quantum}, and extreme nonlinear optics in quantum realm (e.g., high-harmonic generation using quantum light)~\cite{gorlach2020quantum, gorlach2023high, pizzi2023light}. 
Since photons do not interact directly with each other, it is challenging to produce such correlations of photons.
One method widely used is to employ the nonlinear optical response of a medium (such as parametric down-conversion) to produce correlated quantum light~\cite{Drummond01021981, wu1986generation, sanchez2021squeezed}.
However, nonlinear interaction is usually weak.
In the microwave regime, light with strong quantum correlations can be produced using the ultra-strong coupling between photons and superconducting circuits~\cite{wallraff2004strong, hofheinz2009synthesizing}, which, however, is not scalable to a large number of microwave photons.
It was proposed that two-photon emission from independent multilevel atoms can produce squeezed quantum light if the one-photon processes can be suppressed~\cite{scully1985correlated, gauthier1992realization}, but many-body correlations are absent in such independent multilevel emitters.
Bose-Einstein condensation of photons can emerge in thermalized media, but such photons obey the grand canonical statistics and do not exhibit quantum correlations~\cite{schmitt2014observation}. 
Mirrorless superradiant lasing from atoms with coherent interaction is predicted recently~\cite{Bychek2025}, but there the dipole-dipole exchange provides an effective cavity rather than many-body correlations.

To address the challenge of producing correlated many-photon states, here we propose the idea of quantum many-body lasers (QMBLs, Fig.~\ref{fig:figure1}c), in which the interaction-induced correlations in a quantum many-body emitter (QMBE) is transferred to photons through superradiant lasing.
Many-body correlations are ubiquitous in interacting quantum systems.
Under certain conditions, such correlations in `matter' can be transferred to light.
For example, in exciton-polariton Bose-Einstein condensation, the coherence of excitons is transferred to photons~\cite{deng2002condensation}.
Recent advances in quantum technology have made coherently interacting emitters available, including trapped ions~\cite{franke2023quantum}, ultra-cold atoms~\cite{eckner2023realizing}, Rydberg atoms~\cite{bornet2023scalable}, and spins in solids~\cite{choi2017observation, wu2025spin}.
Superradiant transients have been observed in pumped ultracold quantum gases~\cite{baumann2010dicke, zhang2021observation}
and solid-state spin systems~\cite{angerer2018superradiant}.
These systems are promising candidates of QMBEs for realizing QMBLs.

To prove the concept of QMBLs, we consider the Lipkin-Meshkov-Glick (LMG) model as a QMBE (Fig.~\ref{fig:figure1}c). 
In this model, the all-to-all one-axis twisting interaction between many spin-1/2's causes quantum squeezing of the spins~\cite{kitagawa1993squeezed}.
Through the collective interaction between the spins and the cavity field, the phase of the cavity photons becomes locked to the synchronized spin excitations, i.e., magnons.
Therefore, the quantum squeezing within the spins is preserved and transferred to the light during collective emission, leading to a squeezed superradiant laser. 
This superradiant QMBL offers a new scheme for producing bright quantum light.

The system is composed of $N$ interacting spin-1/2's and a cavity.
The Hamiltonian of the system in the rotating reference frame of the cavity field reads (with $\hbar$ set as unity)
\begin{align}
\hat{H}= & i{g}\sum_{j=1}^{N}\left(\hat{a}^{\dagger}\hat{s}_{j}^{-}-{\rm H.c.}\right)+\Delta\sum_{j=1}^{N}\hat{s}_{j}^{z}-\frac{\epsilon}{N}\sum_{i, j=1}^N\hat{s}_{i}^{z}\hat{s}_{j}^{z}, \label{eq:LMGmodel}
\end{align}
where $\hat{a}^{\dagger}$ ($\hat{a}$) is the creation (annihilation) operator of the quantized cavity field, $\hat{s}_{j}^{z}$ and $\hat{s}_{j}^{\pm}$ are the $z$-component and the raiser/lower operators of the $j$-th spin, respectively, and $\Delta$ denotes the detuning between the spin transition and the cavity field. 
Considering that all spins are identically coupled to the cavity and to each other, we can rewrite the Hamiltonian in terms of collective spin operators $\hat{J}^{\pm/z}=\sum_{j=1}^{N}\hat{s}_{j}^{\pm/z}$. The spin-photon interaction becomes $\hat{H}_{\rm int}=i{g}\left(\hat{a}^\dagger\hat{J}^--\hat{a}\hat{J}^+\right)$ with the coupling strength ${g}$.
The large spin radiates photons collectively, with intensity scaling as $N^2$ and phase locked to the magnon oscillation, leading to superradiant lasing~\cite{bonifacio1971quantum}.
The spin-spin interaction can be written as $\hat{H}_{0}=\Delta\hat{J}^{z}-\frac{\epsilon}{N}\left(\hat{J}^{z}\right)^{2}$.
This one-axis twisting interaction induces self-phase-modulation and hence quantum squeezing among the spins~\cite{kitagawa1993squeezed}. The spins are incoherently pumped (via auxiliary energy levels not included in the model). The evolution of the system can be described by the master equation
\begin{equation}
\frac{d\hat{\rho}}{dt}=-i\left[\hat{H}, \hat{\rho}\right]-\frac{\kappa}{2}\mathcal{L}_{\hat{a}}\hat{\rho}-\frac{\gamma}{2}\sum_{j=1}^N\mathcal{L}_{\hat{s}_j^-}\hat{\rho}-\frac{w}{2}\sum_{j=1}^N\mathcal{L}_{\hat{s}_j^+}\hat{\rho}-\gamma_\phi\sum_{j=1}^N\mathcal{L}_{\hat{s}_j^z}\hat{\rho}, 
\label{eq:ME}
\end{equation}
where the dissipation processes are modeled by Lindblad terms as $\mathcal{L}_{\hat{O}}\hat{\rho}=\hat{O}^{\dagger}\hat{O}\hat{\rho}+\hat{\rho}\hat{O}^{\dagger}\hat{O}-2\hat{O}\hat{\rho}\hat{O}^{\dagger}$ for cavity loss with a leakage rate $\kappa$, spontaneous emission of individual spins into free space with an emission rate $\gamma$, incoherent pumping of spins with a pump rate $w$, and spin dephasing rate $\gamma_\phi$.

\begin{figure}[t]
\includegraphics[width=0.9\textwidth]{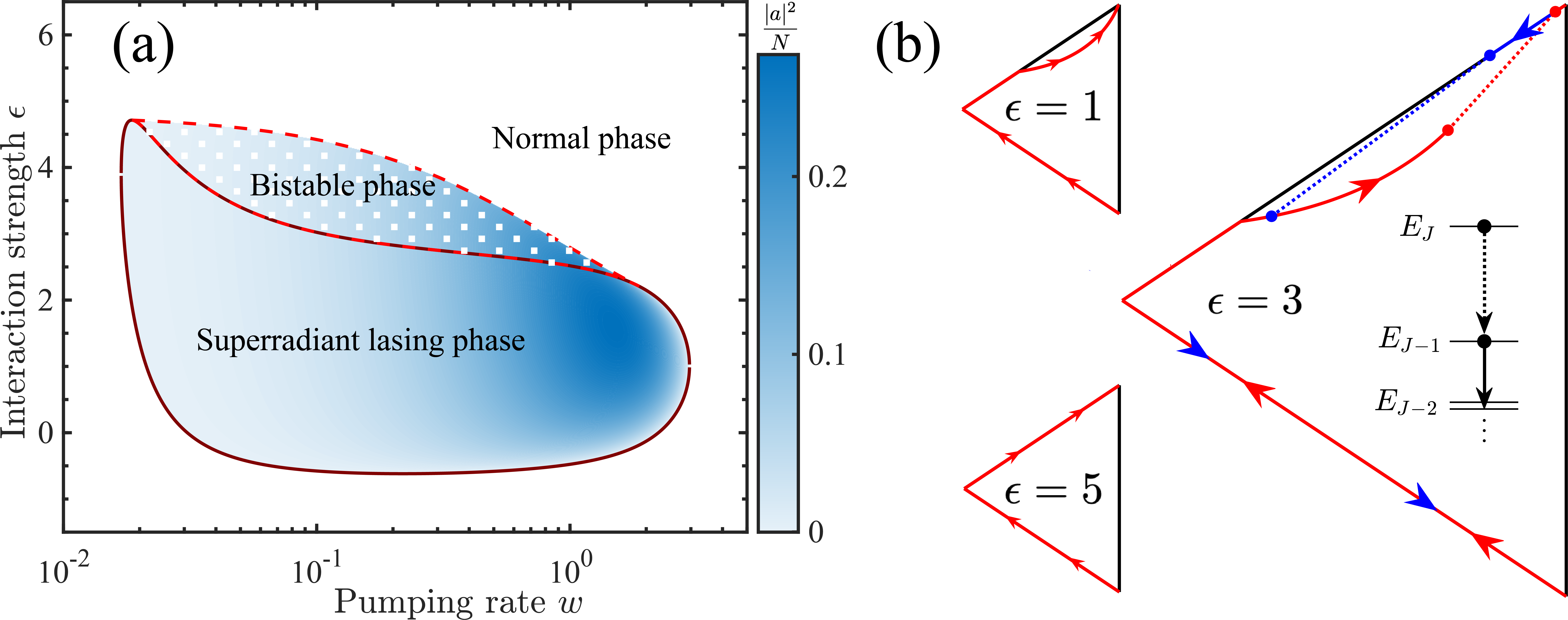}
\caption{\label{fig:figure2}
Superradiant lasing of a pumped cavity-spin system with varying pump rate and spin-spin interaction strength. 
(a) Phase diagram shown by the number of photons in the cavity $\left|{{}} a{{{}}}\right|^2$ normalized by the number of spins $N$, including: I. normal phase, where $\left|{{}} a{{{}}}\right|^2/N=0$  for $N\rightarrow\infty$, II. superradiant lasing phase,  where $\left|{{}} a{{{}}}\right|^2/N\sim O\left(1\right)$, and III. bistable phase. 
Parameters are such that $\gamma=0.01$ and $\gamma_\phi=\kappa=g\sqrt{N}=\Delta=1$.
(b) The stationary total spin $J_{\rm MF}$ (vertical) and its $z$-component ${{}} J^z{{{}}}$ with increasing/decreasing (red/blue arrows) for weak, strong, and moderate spin-spin interaction. The red/blue dotted line in the diagram for $\epsilon=3$ indicates the jump from the lasing/normal state to the normal/lasing state in the bistable phase.
Inset shows the polarization-dependent blockade effect for moderate spin-spin interaction, that is, due to the mean-field energy, the spin transition is far off-resonant from the cavity mode when the polarization is large and become near resonant with reducing the polarization.
}
\end{figure}

According to the mean-field theoretic analysis~\cite{Sup}, lasing occurs when the effective cooperativity of the system is above the threshold, i.e., 
\begin{equation}
    C\equiv C_0\frac{w-\gamma}{w+\gamma} \frac{\Gamma^2}{\Gamma^2+4\Delta_{\epsilon}^2} >1, 
    \label{eq:threshold}
\end{equation} 
where $C_0\equiv {4N{g}^2}/\left({\kappa\kappa_{\rm s}}\right)$ is the intrinsic cooperativity, the factor $(w-\gamma)/(w+\gamma)$ takes into account the population effect, $\kappa_{\rm s}\equiv w+\gamma+\gamma_\phi$ is the decay rate of the magnon mode coupled to the cavity,  $\Gamma\equiv \kappa_{\rm s}+\kappa$ is the total dissipation rate, and the Lorentzian shape accounts for the reduction due to the detuning between the cavity mode and the spin transition
$$\Delta_\epsilon\equiv\Delta-\frac{2\epsilon}{N}{{}} J^z{{{}}}, $$ in which the second term is the mean-field correction of the frequency of the spin transition with $J^z$ being the expectation value of the total spin along the $z$-axis. With varying the pump rate and the spin interaction strength, as shown in the phase diagram in Fig.~\ref{fig:figure2}a, the system can be in a non-lasing normal phase, a superradiant lasing phase, or a bistable phase. In the bistable phase, the system remains in the superradiant lasing state (the normal state) when the increasing (decreasing) pump rate enters the phase boundary, and a first-order transition to the normal state (the superradiant lasing state) occurs when the increasing (decreasing) pump rate exits the boundary.

The phase diagram can be understood by examining the collective spin states under different interaction and pumping conditions.
In the mean-field approximation, up to small fluctuations $\sim\sqrt{N}$, the total spin and its $z$-component have definite stationary values, $J_{\rm MF} \in \{N/2, N/2-1, \ldots, 0\}$ and ${{}} J^z{{{}}} \in\left[-J_{\rm MF}, J_{\rm MF}\right]$, respectively. As shown in Fig.~\ref{fig:figure2}b, the total spin and its $z$-component evolve with increasing or decreasing pump rate $w$ along different trajectories for different spin-spin interaction strengths.

In the case of weak spin-spin interaction ($\left|\epsilon\right|/C_0\ll \Gamma$), with increasing pump rate, the system undergoes a second order phase transition between the normal phase and the superradiant lasing phase as described by the established superradiant laser theory~\cite{carmichael2013statistical}. 
When the spins are under-pumped ($w\lesssim\gamma$), the cooperativity is negative or small ($C< 1$). The collective spin is in the polarized state with ${{}} {J}^{z}{{{}}}=\frac{N}{2}\frac{w-\gamma}{w+\gamma}\approx -J_{\rm MF}$ (see the weak interaction case in Fig.~\ref{fig:figure2}b), making collective emission and hence lasing impossible. With increasing pump rate such that $C>1$, the spins will be pumped into a superradiant state with $1 \ll {{}} J^z{{{}}} < J_{\rm MF}$. Therefore, superadiant lasing is achieved with both optical and magnon coherence, as quantified by 
\begin{subequations}
\begin{eqnarray}
  {{}} {J}^{z}{{{}}}  & = & \frac{N}{2}\frac{w-\gamma}{w+\gamma}C^{-1}, \\
  \left\vert{{}} {J}^{-}{{{}}}\right\vert & = & \left|{{}} {J}^{z}{{{}}} \right| \sqrt{\frac{\left(w+\gamma\right)\left(C-1\right)}{\kappa_{\rm s}/2}},\\
\left\vert{{}} {a}{{{}}}\right\vert^2 & = &    \left|{{}} {J}^{z}{{{}}} \right| \frac{w+\gamma}{\kappa}\left(C-1\right),
\end{eqnarray}
\label{eq:MESP}
\end{subequations}
where $J_z\equiv \left\langle \hat{J}^z (t)\right\rangle$, $J^-\equiv \left\langle \hat{J}^-(t)\right\rangle$, and $a\equiv\left\langle \hat{a}(t)\right\rangle$ are the stationary values for $t\rightarrow \infty$.
 Further increasing the pump rate would over-pump the spins to the polarized state with ${{}} J^z{{{}}} \approx J_{\rm MF}$, which prevents superradiance and hence lasing.

For strong spin-spin interaction ($\left|\epsilon\right|/C_0\gg \Gamma$), the mean-field energy correction $-2{\epsilon}{{}} J^z{{{}}} / N$ would induce a large detuning ($\Delta_\epsilon\gg\Gamma$) between the spin transition and the cavity mode unless ${{}} J^z{{{}}} /N \approx 0 $, preventing superradiant emission into the cavity mode. 
Therefore, with increasing pump rate, the total spin polarization ${{}} J^z{{{}}}$ increases from $-N/2$ to $N/2$ but the collective spin is always in the fully polarized  state with $J_{\rm MF}=\left|{{}} J^z{{{}}}\right|$ (see the strong-interaction case in Fig.~\ref{fig:figure2}b), without superradiant lasing or magnon coherence.

The bistable phase occurs when the interaction is intermediate. Physically, the bistability is due to the dependence of the detuning $\Delta_{\epsilon}$ on the spin polarization ${{}} J^z{{{}}}$ (see the inset of the moderate-interaction case in Fig.~\ref{fig:figure2}b), which leads to a self-consistent dependence of the cooperativity on the spin polarization. 
When the pump rate $w$ increases from $0$ (the red trajectory of the moderate-interaction case in Fig.~\ref{fig:figure2}b), the spins would first be under-pumped with $\left|{{}} J^z{{{}}}\right|\approx J_{\rm MF}$, then for sufficiently strong pump enter the superradiant lasing state with $1\ll {{}} J^z{{{}}} < J_{\rm MF}$ where magnon-photon resonance is reached ({$\Delta_{\epsilon}\lesssim \Gamma$}), and finally become over-pumped with ${{}} J^z{{{}}} \approx J_{\rm MF}$ and off-resonant with the cavity mode.
However, when the pump rate decreases from the over-pumped side (blue trajectory of the moderate-interaction case in Fig.~\ref{fig:figure2}b), the detuning suppresses the emission into the cavity mode (the interaction-induced blockade effect) and therefore the spins maintain the polarization ${{}} J^z{{{}}}\approx J_{\rm MF}$ for lower pump rates than along the rising trajectory. A first-order transition into the superradiant lasing phase occurs when the pump rate is too low to maintain the large spin polarization.

We also perform an exact numerical solution of Eq.~\eqref{eq:ME} for a finite system of $N=15$ spins to validate the mean-field approximation~\cite{Sup}. 
The phase diagram extracted from the qualitative features of the photon-number distribution resembles that obtained by the mean-field theory~\cite{Sup}, with discrepancies attributable to finite-size effects. 

Now we show that the spin squeezing effect can be transferred to the photons when superradiant lasing occurs. 
The one-axis-twisting interaction $\hat{H}_0$ is known to nonlinearly modulate the phase of a spin coherent state, leading to spin squeezing~\cite{kitagawa1993squeezed}. 
In the superradiant lasing phase, the spins and photons are driven into coherent states whose relative phases are locked with only slow diffusion. Therefore, we expect that the phase modulation and the squeezing of the spins are transferred to the photons.
To analyze the fluctuations and correlations of the photons, we employ the Fokker-Planck equation~\cite{carmichael2013statistical}. Considering that the amplitude fluctuations of the spins and photons and their relative phase fluctuations are small quantities, we derive an effective linearized Fokker-Planck equation for the photons~\cite{Sup}
\begin{equation}
    \partial_t \tilde{P}=\left[{\kappa_a}\partial_z z+\frac{1}{2} D_a\partial_z^2 + \frac{1}{2} D_{\phi} \partial^2_{\phi} + \frac{ {\kappa_a}\chi}{\left|{{}} a{{{}}}\right|}  \left( 2z-\frac{1}{2}\partial_z\right)\partial_{\phi}\right]\tilde{P}, 
\label{eq:effFP}
\end{equation}
where $\tilde{P}(z, \phi, t)$ is the $P$-representation of the photon state with $z$ and $\phi$ being the deviations of the photon amplitude and phase from the mean-field value ${{}} a{{{}}}$, respectively.
The drift and diffusion rates of the photon amplitude are, respectively,
\begin{subequations}
\begin{align}
   {\kappa_a} & \equiv \kappa\left(1-C^{-1}\right), \\
    D_a & \equiv \kappa C^{-1}\left[\frac{1}{4}\frac{w+\gamma}{w-\gamma}+\frac{\kappa_{{\rm s}}+w+\gamma}{8C\kappa_{{\rm s}}}+\frac{\left(C-1\right)\left(w+\gamma\right)}{2C\kappa_{{\rm s}}}\right],
    \end{align}
\end{subequations}
and the diffusion rate of the photon phase is
\begin{equation}
D_\phi \equiv \frac{\kappa C}{\left|{{}} a{{{}}}\right|^{2}}\frac{\kappa_{{\rm s}}^{2}}{\Gamma^{2}}\left[\frac{1}{4}\frac{w+\gamma}{w-\gamma}+\frac{\kappa_{{\rm s}}+w+\gamma}{8C\kappa_{{\rm s}}}\right].
\end{equation}
The phase-diffusion term sets the linewidth of the laser, which approaches $\kappa_{\rm s}/N$  and  $\kappa/\left|{{}} a{{{}}}\right|^2$ in the bad  ($\kappa \gg \kappa_{\rm s}$) and good  ($\kappa \ll \kappa_{\rm s}$) cavity limits, respectively [under the condition that $C\sim \mathcal{O}(1)$].
The spin-spin interaction induces an amplitude-dependent drift of the photon phase, with a rate $\chi {\kappa_a}/\left|{{}} a{{{}}}\right|$, where 
\begin{equation}
    \chi\equiv \frac{{ 2}{{}} J^z{{{}}}}{N} \frac{ \epsilon}{ \Gamma}
\end{equation}
is the mean-field interaction normalized by the total dissipation rate.
To understand the effect of the interaction term, let us consider an initial state $\tilde{P}(z, \phi, 0)= e^{-z^2 {\kappa_a} /D_a }f(\phi)$ (which is stationary under the drift and diffusion of amplitude). On this state the $\chi$-related term in Eq.~\eqref{eq:effFP} induces a positive phase drift for $z>0$ and a negative phase drift for $z<0$, that is, twisting of the photon quasi-probability distribution.
Physically, the twisting of the photon state is a combined effect of the many-body interaction in the spins and superradiant lasing. First, due to the spin interaction, the frequency of the magnon oscillation depends on the spin polarization ${{}} J^z{{{}}}$ (through the mean field $\chi$); second, the frequency-dragging effect of superradiant lasing~\cite{bohnet2012steady} leads to the dependence of the photon frequency on the spin polarization and, in turn, on the photon amplitude $z$.

\begin{figure}[t]
\includegraphics[width=0.8\textwidth]{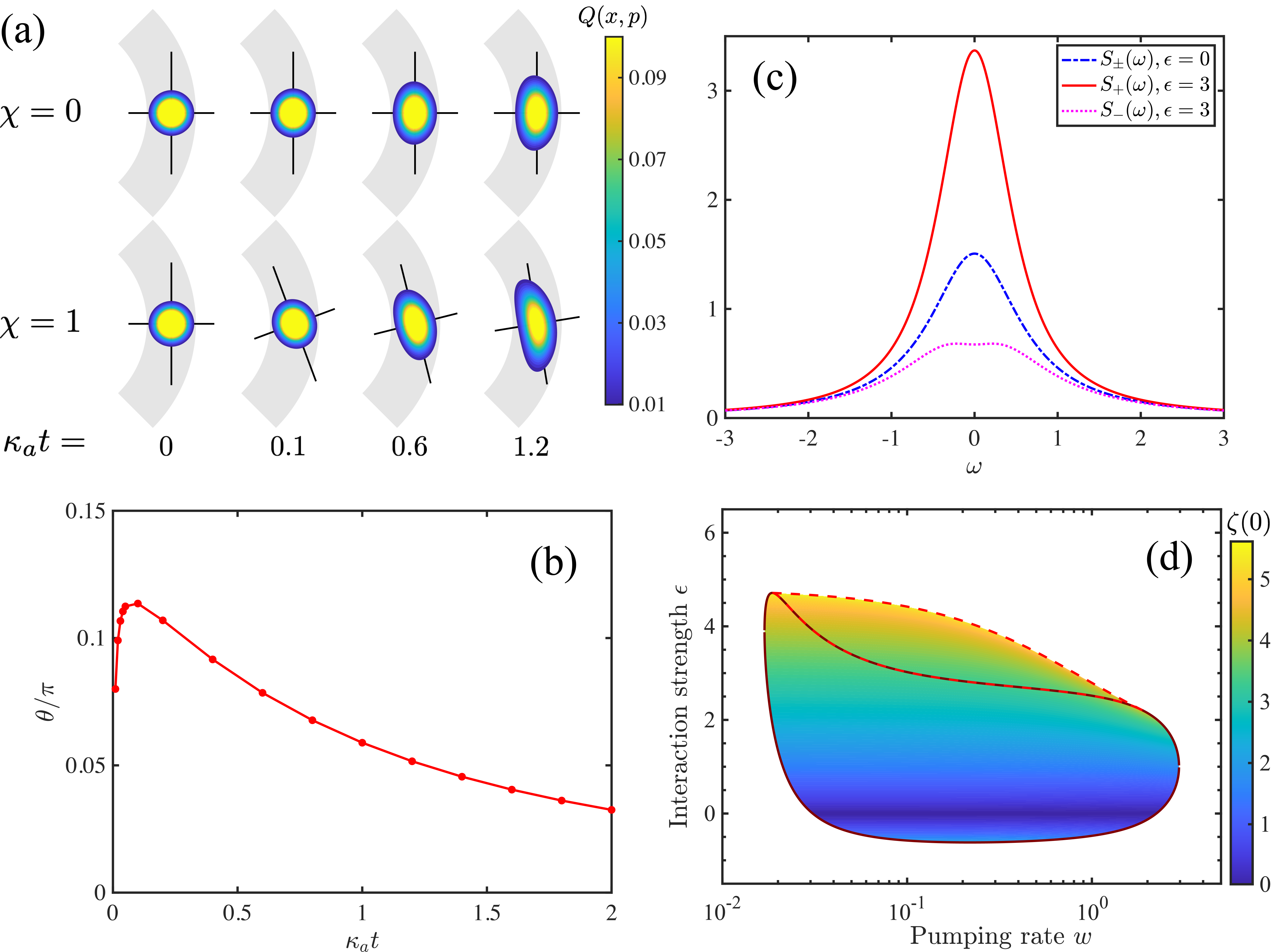}
\caption{\label{fig:figure3}
Squeezing in superradiant lasing from spins with all-to-all interactions. (a) Evolution of Husimi Q-representation in the phase space from an initial coherent state for spin interaction strength  $\chi=0$ (upper row) and $\chi=1$ (lower row). The lines are the principal axes of the quadrature fluctuations at zero frequency.  (b) The angle of the axis along which the zero-frequency noise is mimimum [i.e., the axis for $S_-(0)$]. Here, we set $\kappa_a=1,D_a=0.1,D_\phi=0.01$, and $\vert {{}} a{{{}}}\vert^2=100$ to illustrate the twisting effect.
(c) Principal noise spectra $S_{\pm}(\omega)$ of the laser for $\epsilon=0$  (blue dash-dotted line) and $\epsilon=3$ (red solid and magnet dashed lines). The pump rate is set as $w=0.2$.
(d) Decibel squeeze parameter $\zeta(0)$ as a function of the pump rate and interaction strength. In the bistable phase, only the superradiant lasing state is considered. Parameters not otherwise specified are the same as in Fig.~\ref{fig:figure2}.
}
\end{figure}

To quantify the squeezing effect, we consider the photon dynamics starting from a coherent state $\tilde{P}(z, \phi, t=0)=\delta(z)\delta(\phi)$, which can be understood as a transient state of a stationary laser. The exact solution of Eq.~(\ref{eq:effFP})  clearly shows the twisting and squeezing effect, as shown Fig.~\ref{fig:figure3}a and b, which present the Husimi $Q$-representations, transformed from the $P$-representation but providing a smooth and non-negative visualization in the phase space with the quadratures defined as $\hat{x}\equiv \left(\hat{a}^{\dagger}+\hat{a}\right)/\sqrt{2}$ and $\hat{p}\equiv i\left(\hat{a}^{\dagger}-\hat{a}\right)/\sqrt{2}$.
This evolution from the transient coherent state for a short time ($t\ll D^{-1}_{\phi}$)  can be reproduced by the following master equation for the reduced density matrix of the photons
 \begin{equation}
            \frac{d\hat{\rho}_{\rm ph}}{dt}= -i\left[\hat{H}_{\rm eff} , \hat{\rho}_{\rm ph}\right] -\left[\kappa_a{\mathcal L}_{\hat{a}}+D_a\mathcal{L}_{\hat{p}}+{\frac{1}{2}D_{\phi}}{\mathcal L}_{\hat{a}^{\dagger}\hat{a}}\right]\hat{\rho}_{\rm ph}, 
\label{eq:effMasterEq}
 \end{equation}
with an effective Hamiltonian  $\hat{H}_{\rm eff}=\sqrt{2}\kappa_a \left|{{}} a{{{}}}\right| \hat{p}+\frac{1}{2}{\left|{{}} a{{{}}}\right|^{-2}}{\kappa_a} {\chi}\left(\hat{a}^{\dagger}\hat{a}-\left|{{}}  a{{{}}}\right|^{2}\right)^2$, which, for the initial coherent state, contains a squeezing term $\approx \frac{1}{2} \chi \kappa_a \left( \hat{a}^{\dagger}\hat{a}^{\dagger}+\hat{a}\hat{a}\right)$ resulting from the spin-spin interaction.

We solve Eq.~\eqref{eq:effMasterEq} to calculate the squeeze parameter. The noise spectra $S_{\pm}(\omega)$ along the two principal axes are obtained by diagonalizing the real part of the Fourier transform of the symmetric quadrature correlation matrix composed of $\left\{ \delta \hat{x}(t),\delta \hat{x}(0) \right\}/2$, $\left\{ \delta \hat{x}(t),\delta \hat{p}(0) \right\}/2$, $\left\{ \delta \hat{p}(t),\delta \hat{x}(0) \right\}/2$, and $\left\{ \delta \hat{p}(t),\delta \hat{p}(0) \right\}/2$.
As shown in Fig.~\ref{fig:figure3}c, there is no squeezing [$S_+(\omega)=S_-(\omega)$] when the spin-spin interaction is absent ($\epsilon=0$). 
When the spins interact ($\epsilon \neq 0$), the two principal noise spectra are unequal, with one reduced from the coherent state, which is a signature of squeezing.
The decibel squeeze parameter at zero frequency ($\omega=0$, for squeezing in the stationary state) is obtained as~\cite{Sup}
\begin{equation}
    \zeta(0)\equiv -10 \log_{10} \frac{S_-(0)}{\left. S_{-}(0)\right|_{\epsilon=0}}=\frac{20}{\ln10}\operatorname{arcsinh}(\vert\chi\vert), 
\label{eq:Squ}
\end{equation}
as presented in Fig.~\ref{fig:figure3}d. 
For strong interactions ($\vert\chi\vert\gg 1$), $\zeta(0)\approx 20 \log_{10}(2\vert\chi\vert)$.
Given a spin-cavity system with a large intrinsic cooperativity $C_0$, strong spin-spin interaction $\epsilon\gg \kappa_{\rm s}+\kappa$, and a detuning compensating the mean-field correction $\Delta\lesssim \epsilon$, a bright laser with large squeezing can be realized.

The realization of squeezed superradiant lasing in a quantum many-body emitter or, more generally, a QMBL requires two key elements, namely, superradiant coupling (characterized by a large intrinsic cooperativity) and coherent many-body interaction (with mean-field interaction strength greater than or at least comparable to the total dissipation rate).
Several experimental platforms currently available in laboratories can be adopted for realizing the QMBL, such as Rydberg atoms~\cite{bornet2023scalable}, cold atoms~\cite{norcia2018cavity}, and  spins in solids~\cite{wu2025spin}.
For example, we consider nitrogen-vacancy (NV) center spins in diamond as a candidate system, where both superradiance~\cite{angerer2018superradiant} and masing~\cite{breeze2018continuous} have been demonstrated.
In such a system, the total dissipation of magnons and microwave photons includes the spin relaxation (with a typical rate $\gamma\sim 2\pi\times30$~Hz for $T_1\sim5$~ms), the microwave cavity leakage (with a typical rate $\kappa\sim 2\pi\times1$~MHz for a cavity with a quality factor $Q\sim10^4$ at the masing frequency of $2\pi\times10$~GHz), the spin dephasing (with a rate $\gamma_\phi=2/T_2^{*}$ depending on the densities of background electron and nuclear spins), and the incoherent pumping. We can choose a diamond with a density of {nitrogen}  $\sim 10$~ppm such that $\gamma_\phi\sim 2\pi\times1$~MHz (for $T_2^*\sim 0.3$~$\mu$s)~\cite{Zhao2012}. The direct spin-spin interaction is too weak to realize the condition of coherent interaction, that is, the mean-field interaction energy $\epsilon\ll \Gamma = \kappa+\gamma+w+\gamma_\phi\sim 2\pi\times2$~MHz. Instead, we can adopt spin-spin interaction mediated by an off-resonant auxiliary cavity mode~\cite{norcia2018cavity}. For a detuning between the spin transition and the auxiliary cavity mode $\Delta_{\rm A}\sim 2\pi\times20$~MHz, coupling between the  auxiliary mode and a single spin ${g}_{\rm A}\sim 2\pi\times0.2$~Hz, and the number of NV spins resonantly coupled to the masing cavity mode $N\sim 2\times 10^{15}$ (estimated with a typical nitrogen-to-NV conversion ratio $\sim 20\%$, a charge state ratio of NV$^-$ to NV$^0$ $\sim 70:30$,  $1/8$ NV$^{-1}$'s chosen from the four crystallographic orientations and the two $^{15}$N nuclear spin states, and a  diamond size $\sim 4\times4\times 4$~mm$^3$), the coupling strength is estimated as high as $\epsilon\simeq N{g}_{\rm A}^2/\Delta_{\rm A}\sim 2\pi\times4$~MHz~$>\Gamma$ with the auxiliary cavity leakage rate $\kappa_{\rm A}\sim2\pi\times1$~MHz.
When the effective detuning is negligible ($\Delta_\epsilon\approx0$) and the coupling between the masing cavity mode and a single spin is ${g}\sim 2\pi\times0.02$~Hz, we estimate an effective cooperativity of $C\approx C_0\sim 3$ (above the lasing threshold) for $w\sim 2\pi\times0.8$~kHz, and the squeezed superradiant maser would have an output power $P_{\rm out}\sim 3$~$\rm \mu$W, a linewidth $D_\phi/( 2\pi) \sim 0.6$~mHz, and a squeeze parameter $\zeta(0)\sim 5$~dB (with $\chi \sim 0.6$).

In summary, we put forward the concept of quantum many-body laser, where superradiant lasing from many pumped emitters that have coherent interactions can transfer many-body correlations among the emitters to photons. As a demonstration of the idea, we investigate superradiant lasing from a spin systems with all-to-all interaction and show that the spin squeezing induced by the coherent interaction can be transferred to photons through superradiant lasing. 
Several experimental platforms such as spins in solids and superradiant Rydberg atoms and cold atoms can be candidates for realizing quantum many-body lasers. It is conceivable that the QMBL can also transfer other types of higher-order quantum correlations beyond squeezing from quantum many-body emitters to photons. QMBL, being bright light sources with non-classical correlations, can find applications in quantum technologies and strong-field quantum optics, including quantum metrology, extreme nonlinear quantum optics, and quantum communication. 

\begin{acknowledgments}
    This work was supported by the National Natural Science Foundation of China/Hong Kong Research Council Collaborative Research Scheme Project CRS-CUHK401/22, the New Cornerstone Science Foundation, and the Hong Kong Research Grants Council Senior Research Fellow Scheme Project SRFS2223-4S01.
\end{acknowledgments}

\appendix

\section{Mean-field analysis}
In the mean-field theory, we neglect the quantum fluctuations since they are much smaller than their corresponding mean-field contributions. Thus we derive from Eq.~(\ref{eq:ME}) in main text the mean-field equations in the reference frame rotating at the laser frequency $\omega_{\rm L}$,
\begin{subequations}
\label{eq:MFT_eq}
\begin{align}
 & \frac{d {a}\left(t\right) }{dt}=\left[-i\left(\omega_{\rm c}-\omega_{\rm L}\right)-\frac{\kappa}{2}\right] {a}\left(t\right) +{g} {J}^{-} \left(t\right),\\
 & \frac{d {J}^{-}\left(t\right) }{dt}= \left[-i\left(\omega_{\rm c}+\Delta_{\epsilon}-\omega_{\rm L}\right)-\frac{\kappa_{\rm s}}{2} \right] {J}^{-} \left(t\right)
 +2{g} {J}^{z}\left(t\right)  {a}\left(t\right) ,\\
     & \frac{d {J}^{z}\left(t\right) }{dt}=\frac{N}{2}\left(w-\gamma\right)-\left(w+\gamma\right) {J}^{z}\left(t\right) -{g}  \left[ {a} \left(t\right) {J}^{+}\left(t\right)
  + {\rm c.c.} \right],\label{eq:MFeq}
\end{align}
\end{subequations}
where $\omega_{\rm c}$ is the cavity frequency, $J^+\left(t\right)=\left[J^-\left(t\right)\right]^*$, and the spin-cavity coupling $g$ is assumed to be a real number.

\begin{figure}[htbp]
\includegraphics[scale=0.4]{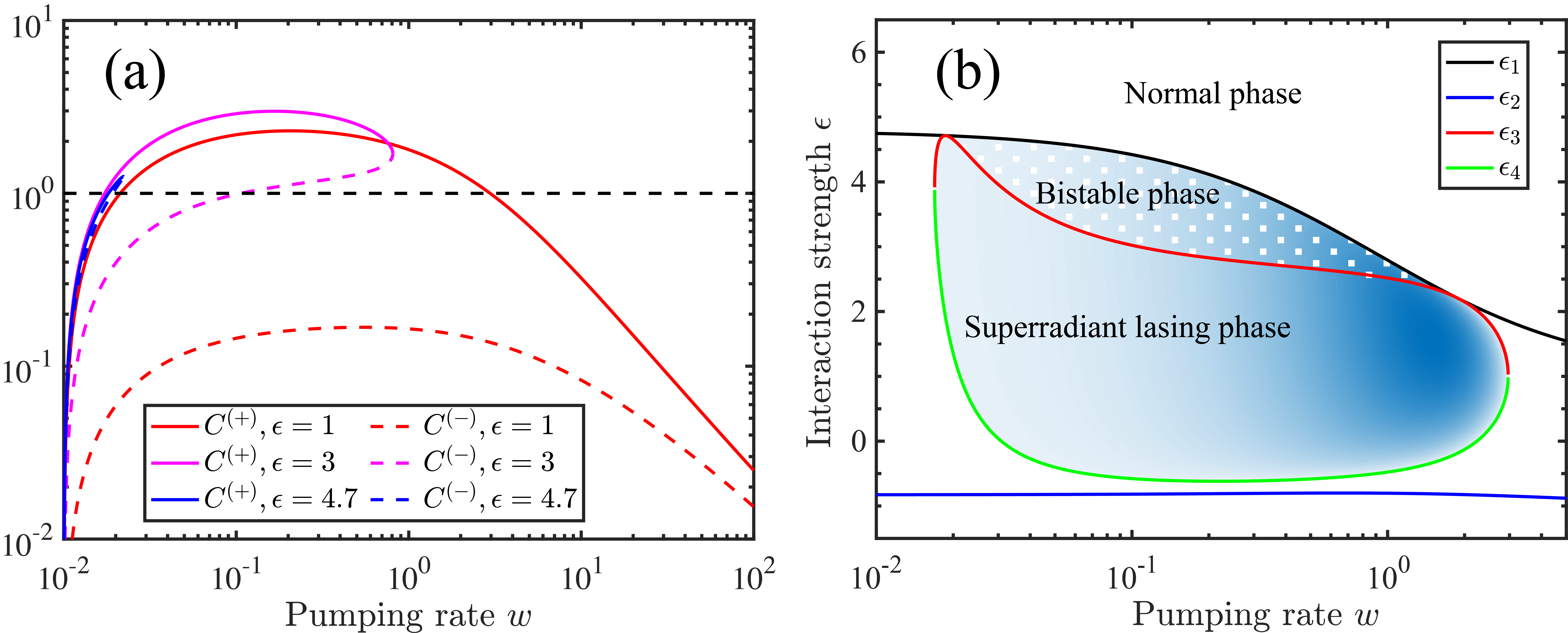}
\caption{\label{fig:figureS1}
(a) Effective cooperativities ${C}^{(\pm)}$ versus pump rate $w$ for different interaction strengths: $\epsilon=1$ (red), $\epsilon=3$ (magenta), and $\epsilon=4.7$ (blue). Solid/dashed curves denote ${C}^{(+/-)}$. The black dashed line indicates the lasing threshold $C=1$.
(b) Phase boundaries $\epsilon_{1/2/3/4}\left(w\right)$ as functions the pump rate $w$. The normal phase, the superradiant lasing phase, and the bistable phase occur correspondingly for $1>{C}^{(+)}>{C}^{(-)}$, ${C}^{(+)}>1>{C}^{(-)}$, and ${C}^{(+)}>{C}^{(-)}>1$.}
\end{figure}

The stationary solutions of the cavity field amplitude $a$, spin polarization $J^z$, and magnon amplitude $J^{\pm}$ can be obtained by setting the time-derivatives in Eq.~\eqref{eq:MFT_eq} to be zero. Consequently, the laser frequency is obtained as $$\omega_{\rm L}=\omega_{\rm c}+\frac{\kappa}{\Gamma}\Delta_\epsilon.$$
The steady-state solutions of the mean-field equations are governed by the lasing threshold condition for the effective cooperativity ${C}>1$. Figure \ref{fig:figureS1}a shows ${C}$ as a function of the pump rate $w$ for several interaction strengths $\epsilon$.
Because ${C}$ is determined by a nonlinear self-consistent equation of $ J^{z}$, it admits two solutions
\begin{equation}
    C^{(\pm)}=C_{0}\frac{w-\gamma}{w+\gamma}\frac{\Gamma^{2}+8\frac{\epsilon}{C_{0}}\Delta\pm\Gamma\sqrt{\Gamma^{2}+16\frac{\epsilon}{C_{0}}\Delta-16\frac{\epsilon^{2}}{C_{0}^{2}}}}{2\left(\Gamma^{2}+4\Delta^{2}\right)}>1,
\end{equation}
where $C_0\equiv 4Ng^2/\left(\kappa\kappa_{\rm s}\right)$ is the intrinsic cooperativity.
In the weak-interaction limit ($\epsilon/C_0\ll \Gamma$), we find that ${C}^{(-)}=\frac{4\epsilon^2}{C_0\Gamma^2}\frac{w-\gamma}{w+\gamma}+O\left(\epsilon^3\right)$ and ${C}^{(+)}=C_0\frac{w-\gamma}{w+\gamma}\frac{{\Gamma^2}}{\Gamma^2+{4\Delta^2}}+O\left(\epsilon\right)$, so only the lasing threshold condition ${C}^{(+)}>1$ can be satisfied. This lasing threshold condition is consistent with the established superradiant laser theory~\cite{carmichael2013statistical}.
In the strong-interaction limit ($\epsilon\rightarrow\infty$), ${C}^{(\pm)}$ have no real-number solutions, so the lasing threshold condition cannot be satisfied.
In the intermediate regime, where the nonlinearity is strong but does not completely suppress superradiant lasing, both ${C}^{(\pm)}$ can satisfy the lasing threshold condition, which leads to the bistability.

The stability of the steady-state solutions is assessed by linearizing the mean-field equations around each steady state. A steady state is stable if and only if all eigenvalues of the Jacobian matrix
\begin{equation}
\mathbb{M}=\left(\begin{array}{ccccc}
-\frac{\kappa}{2}+i\frac{ \kappa \Delta_\epsilon}{\Gamma} & 0 & {g} & 0 & 0\\
0 & -\frac{\kappa}{2}-i\frac{\kappa \Delta_\epsilon}{\Gamma} & 0 & {g} & 0\\
2{g} J^{z}  & 0 & -\frac{\kappa_{\rm s}}{2}-i\frac{\kappa_{\rm s} \Delta_\epsilon}{\Gamma} & 0 & i\frac{2\epsilon}{N} J^{-} +2{g} a\\
0 & 2{g} J^{z} & 0 & -\frac{\kappa_{\rm s}}{2}+i\frac{\kappa_{\rm s} \Delta_\epsilon}{\Gamma} & -i\frac{2\epsilon}{N} J^{+} +2{g} a^{*}\\
-{g} J^{+} & -{g} J^{-} & -{g} a^{*} & -{g} a & -\left(w+\gamma\right)
\end{array}\right)
\end{equation}
have negative real parts. Note that, due to the U(1) symmetry, the phase of $ a$ is arbitrary, the phase of $ a^{*} $ and $ J^{\pm}$ are fixed by the mean-field equations once the phase of $ a$ is chosen. The phase diagram in main text is obtained from this stability analysis. When $1>{C}^{(+)}>{C}^{(-)}$, the only stable steady state is the normal state. When ${C}^{(+)}>1>{C}^{(-)}$, the normal state becomes unstable and the superradiant lasing state corresponding to ${C}^{(+)}$ is stabilized. When ${C}^{(+)}>{C}^{(-)}>1$, the mean-field equations admit two superradiant-lasing solutions but only the branch at ${C}^{(+)}>1$ is stable. In this regime, the normal state also becomes stable. The system is in the bistable phase. The phase boundaries, as shown in Fig.~\ref{fig:figureS1}b, are determined by the conditions $C^{(+)}=C^{(-)}$ and $C^{(\pm)}=1$, which yield the four solutions of interaction strength as functions of the pump rate,
\begin{subequations}
\begin{align}
    C^{+}=C^{-} \ \Rightarrow \ & \epsilon_{1/2}\left(w\right)  =\frac{C_0}{4}\left( {2\Delta}\pm\sqrt{\Gamma^2+{4\Delta^2}}\right),
    \\
    C^{\pm}=1\  \Rightarrow \ & \epsilon_{3/4}\left(w\right)  =\frac{\Gamma\left(w+\gamma\right)}{2\left(w-\gamma\right)}\left(\frac{2\Delta}{\Gamma}\pm\sqrt{C_0\frac{w-\gamma}{w+\gamma}-1}\right).
    \end{align}
\end{subequations}
The superradiant lasing phase occurs for $\epsilon_4<\epsilon<\epsilon_3$, while the bistable phase occurs for $\epsilon_1<\epsilon<\epsilon_4$ and $\epsilon_2<\epsilon<\epsilon_3$. Otherwise, the system is in the normal phase. The asymmetry of the phase diagram arises from the finite detuning $\Delta>0$; for $\Delta=0$, the diagram becomes symmetric.

\section{Fokker-Planck equation}
We derive the Fokker–Planck equation following Ref.~\cite{carmichael2013statistical}, incorporating the effects of the coherent interaction between the spins. The theory is based on the Glauber–Sudarshan $P$-representation, a quasi-probability description of the system state $\hat{\rho}$, defined as
\begin{align}
P(\alpha,\alpha^*;\zeta,\zeta^*,m)
&= \frac{1}{2\pi^5}\int dA^* dA \int dJ^* dJ \int dM \,
\mathrm{Tr}\!\left[\hat{\rho} e^{iA^*\hat{a}^\dagger} e^{iA\hat{a}}
e^{iJ^*\hat{J}^+} e^{iM\hat{J}^z} e^{iJ\hat{J}^-}\right] \notag\\
&\quad \times e^{-i{\mathcal A}^{*}\alpha^{*}}e^{-i{\mathcal A}\alpha}e^{-i{\mathcal J}^{*}\zeta^{*}}e^{-i{\mathcal M} m}e^{-i{\mathcal J}\zeta}.
\label{eq:B1}
\end{align}

Taking the time derivatives of the $P$-representation and substituting Eq.~(\ref{eq:ME}) for $\hat{\rho}$ in main text, we obtain the corresponding Fokker–Planck equation in the reference frame rotating at the laser frequency as
\begin{align}
& \partial_t P\left(\alpha,\alpha^{*},\zeta,\zeta^{*},m\right) \notag\\
= & i\left(\omega_{\rm c}-\omega_{\rm L}\right)\left(\partial_{\alpha}\alpha-\partial_{\alpha^*}\alpha^{*}\right)P +i\left(\omega_{\rm c}+ \Delta-\omega_{\rm L}\right)\left(\partial_{\zeta}\zeta-\partial_{\zeta^*}\zeta^{*}\right)P \notag\\
 & -{g}\left\{\left[\left(e^{-\partial_m}-1\right)\zeta+2m\partial_{\zeta}-\partial_{\zeta}^2\zeta\right]\alpha+\partial_{\alpha}\zeta\right\}P -i\frac{\epsilon}{N}\left(\partial_{\zeta}\zeta+ 2m\partial_{\zeta}\zeta-\partial_{\zeta}^2\zeta^{2} \right)P + {\rm c.c.}
 \notag\\ & 
 +\frac{\kappa}{2}\left(\partial_{\alpha}\alpha+\partial_{\alpha^*}\alpha^{*}\right)P \notag \\  & 
 +\frac{\gamma}{2}\left[\partial_{\zeta}\zeta+\partial_{\zeta^*}\zeta^{*}+\left(e^{\partial_m}-1\right)\left(N+2m\right)\right]P
 \notag \\  & 
 +\frac{\gamma_\phi}{2}\left[\partial_{\zeta}\zeta+\partial_{\zeta^*}\zeta^{*}+\partial_\zeta\partial_{\zeta^*}e^{-\partial_m}\left(N+2m\right)\right]
 P
 \notag\\
 & +\frac{w}{2}\left[\left(e^{-\partial_m}-1\right)\left(N-2m\right)+\partial_{\zeta}^2\partial_{\zeta^*}^2 e^{\partial_m}\left(N+2m\right)+\left(2 e^{-\partial_m}+2 \partial_{\zeta}\partial_{\zeta^*}-1\right)\left(\partial_{\zeta}\zeta+\partial_{\zeta^*}\zeta^{*}\right)+2N\partial_{\zeta}\partial_{\zeta^*}\right]P 
 \notag\\
\equiv & {\mathcal L}\left(\alpha,\alpha^{*},\zeta,\zeta^{*},m,\partial_{\alpha},\partial_{\alpha^*},\partial_{\zeta},\partial_{\zeta^*},\partial_m\right)P\left(\alpha,\alpha^{*},\zeta,\zeta^{*},m\right).
\label{eq:FP0}
\end{align}
This Fokker–Planck equation is not exactly solvable since it contains higher-order derivatives (in $\zeta,\zeta^*$ and $m$) and nonlinear drift/diffusion coefficients.
We perform a linear expansion around the mean-field solutions. However, it is important to note that the mean-field solutions differ qualitatively when the system is in the normal and lasing states; accordingly, the linearization should be carried out relative to the respective mean fields in the normal and lasing states. 

Below the lasing threshold, i.e., for ${C}<1$, keeping drift terms up to linear in $\alpha$, $\zeta$ and $\mu=m- J^z$ (with $ J^z=\frac{N}{2}\frac{w-\gamma}{w+\gamma})$ and diffusion terms up to constant around the normal state, we transform Eq.~(\ref{eq:FP0}) to
\begin{align}
\partial_t P (\alpha,\alpha^*,\zeta,\zeta^*,\mu,t)
\approx &  \left[\frac{\kappa}{2}+{i\left(\omega_{\rm c}-\omega_{\rm L}\right)}\right]{\partial_\alpha}\alpha {P} -{{g}}\zeta  {\partial_\alpha}{P}
+{\rm c.c.} 
\notag \\
+&\left[\frac{\kappa_{\rm s}}{2}+{i\left(\omega_c+\Delta_\epsilon-\omega_{\rm L}\right)}\right]{\partial_\zeta}\zeta P-2 J^z g\alpha {\partial_\zeta}{P}+{\rm c.c.} \notag \\
  + &\left(w+\gamma\right){\partial_\mu}\mu{P}+\frac{N w\gamma}{w+\gamma}\partial^{2}_{\mu}{P}+\frac{Nw\kappa_{\rm s}}{w+\gamma} \partial_\zeta\partial_{\zeta^*}{P},
\label{eq:FP1}
\end{align}
which is Gaussian and numerically solvable. Adiabatically eliminating the spin variables $\zeta$, $\zeta^*$, and $\mu$~\cite{carmichael2013statistical}, we obtain an effective Fokker-Planck equation
\begin{align}
        \partial_t\tilde{P}\left(\alpha,\alpha^*,t\right) & \approx\left\{\left(1-{C}\right)\left[\frac{\kappa}{2}+i\left(\omega_{\rm c}-\omega_{\rm L}\right)\right]\partial_\alpha \alpha \tilde{P} +{\rm c.c.}\right\} +\frac{\kappa w{C}}{w-\gamma}\partial_\alpha\partial_{\alpha^*} \tilde{P},
    \label{eq:FPEff1}
\end{align}
where $\tilde{P}\left(\alpha,\alpha^*,t\right)$ is the quasi-probability distribution for photons.
Note that at the normal state, the spins have no collective coherence and hence the dephasing rate $\gamma_{\phi}$ has no {\em direct} effects on the photon dynamics (except for the reduction of the cooperativity). Equation~(\ref{eq:FPEff1}) can be solved analytically, and we obtain the stationary photon and spin statistics as
\begin{subequations}
\begin{align}
 & \left\langle \hat{a}^{\dagger}\hat{a}\right\rangle =\frac{w{C}}{w-\gamma}\frac{1}{1-{C}}\sim O\left(1\right),
 \\
 & \Delta J^z = \sqrt{ \left\langle\mu^2 \right\rangle }=\sqrt{\frac{Nw\gamma}{\left(w+\gamma\right)^2}}\ll \left| J^z\right|\sim O\left(N\right) ,\\
 & g^{(1)}(\tau)\equiv \frac{\left\langle \hat{a}^{\dagger}\left(t+\tau\right)\hat{a}\left(t\right)\right\rangle }{\left\langle \hat{a}^{\dagger}\hat{a}\right\rangle}=\exp\left\{-\frac{\kappa\tau\left(1-{C}\right)}{2}\left[1-\frac{2i(\omega_{\rm c}-\omega_{\rm L})}{\kappa}\right]\right\},\\
 & g^{(2)}(\tau)\equiv \frac{\left\langle\hat{a}^{\dagger}\left(t\right)\hat{a}^{\dagger}\left(t+\tau\right)\hat{a}\left(t+\tau\right)\hat{a}\left(t\right)\right\rangle }{\left\langle \hat{a}^{\dagger}\hat{a}\right\rangle^2}=1+e^{-{\kappa\tau\left(1-{C}\right)}}.
\end{align}
\label{eq:FP1Sol}
\end{subequations}
The photon statistics is thermal, as indicated by $g^{\left(2\right)}\left(0\right)=2$.
This result coincides with the predictions of traditional superradiant lasing theory~\cite{carmichael2013statistical}, except for small interaction-induced reduction of ${C}$. 

Above the lasing threshold, ${C}>1$, both the magnon mode and the cavity field have macroscopic amplitudes, so their fluctuations can be treated as small quantities. To separate the small fluctuations from the large mean-field values, we define the small amplitude and phase fluctuations as
\begin{subequations}
\begin{align}
z\equiv & \left|\alpha\right| - \left| a\right|, \\
\phi\equiv & \arg\left(\alpha\right) - \Phi, \\
{{r}} \equiv & \left|\zeta\right|- \left|  {J}^{-}\right|, \\
 \varphi \equiv & \arg\left(\zeta\right),\\
\mu\equiv & m-  {J}^{z},
\end{align}
\end{subequations}
where $\Phi= \tan^{-1}\left[{2\left(\omega_{\rm c}-\omega_{\rm L}\right)}/\kappa\right]=-\tan^{-1}\left[{2\left(\omega_{\rm c}+\Delta_{\epsilon}-\omega_{\rm L}\right)/\kappa_{\rm s}}\right]=-\tan^{-1}\left({2\Delta_{\epsilon}}/\Gamma\right)$ is the stationary phase difference between the cavity mode and magnon mode. 
The quasi-probability distribution in $z$, ${{r}}$, $\mu$, $\varphi$, and $\phi$ is defined as
\begin{equation}
\bar{P}\left(z,\phi,{{r}},\varphi,\mu\right)=\left(\left| a\right|+z\right)\left(\left|  {J}^{-}\right|+{{r}}\right)P\left(\alpha,\alpha^{*},\zeta,\zeta^{*},m\right),
\end{equation}
where the factor $\left(\left| a\right|+z\right)\left(\left|  {J}^{-}\right|+{{r}}\right)$ takes into account the conversion of the measure in polar coordinates to that in Cartesian coordinates.
Substituting $\bar{P}$ into the Fokker-Planck equation Eq. (\ref{eq:FP0}) and keeping the drift terms up to linear in $z$, ${{r}}$, $\mu$, $\psi\equiv \phi+\varphi$ and $\delta\equiv \phi-\varphi$ and diffusion terms up to constant, we obtain
\begin{align}
&\nonumber\partial_t\bar{P}\left(z,\psi,{{r}},\delta,\mu\right) \nonumber \\
\approx&\partial_z\left(\frac{\kappa}{2}z-{g}{{r}} \cos\Phi+{g}\left|  {J}^{-}\right|\delta \sin\Phi \right)\bar{P}
 \nonumber \\
+ & \partial_{{r}}\left[\frac{\kappa_{\rm s}}{2}{{r}}-2{g}\left(\left| a\right|\mu+z {J}^{z}\right) \cos\Phi+2{g}\left| a\right| {J}^{z}\delta \sin\Phi\right]\bar{P} 
+ \frac{2Nw+\left(N+2 {J}^{z}\right)\gamma_\phi +4 g \left| a \right| \left| {J}^{-}\right|\cos\Phi }{8} \partial^2_{{r}} \bar{P}
\nonumber \\
+ & \partial_\mu\left[\left(w+\gamma\right)\mu+2{g}\left(\left| a\right|{{r}}+\left|  {J}^{-}\right|z\right) \cos\Phi -2{g}\left| a \right| \left| {J}^{-}\right|\delta \sin\Phi\right]\bar{P}
\nonumber \\ +&
\left(w\frac{N-2  {J}^{z} }{4}+\gamma \frac{N+2  {J}^{z} }{4}-{g}\left| a \right| \left|  {J}^{-}\right| \cos\Phi\right)\partial^2_\mu\bar{P}
\nonumber \\
+&\partial_\psi\left[{g}\left({{{r}}}-\frac{\left|  {J}^{-}\right|z}{\left| a\right|^{2}}-\frac{2 \left| a \right| \left|{J}^{-}\right|\mu+ 2{J}^{z} \left|  {J}^{-}\right|  z- 2{J}^{z} \left| a\right| {{r}}}{\left|  {J}^{-}\right|^2}\right) \sin\Phi+{g}\frac{\left|  {J}^{-}\right|^2-2 {J}^{z} \left| a\right|^2}{\left| a \right| \left|{J}^{-} \right|}\delta \cos\Phi -\frac{2\epsilon}{N}\mu\right]\bar{P}
\nonumber \\ 
+&\partial_\delta\left[
{g}\left({{{r}}}-\frac{\left|  {J}^{-}\right|z}{\left| a\right|^{2}}+\frac{2 \left| a \right| \left|{J}^{-}\right|\mu+ 2{J}^{z} \left|  {J}^{-}\right|  z- 2{J}^{z} \left| a\right| {{r}}}{\left|  {J}^{-}\right|^2}\right)\sin\Phi
+{g}\frac{\left|  {J}^{-}\right|^2+2 {J}^{z} \left| a\right|^2}{\left|a \right| \left| {J}^{-}\right|}\delta \cos\Phi+\frac{2\epsilon}{N}\mu 
\right]\bar{P}
 \nonumber \\
+&\left({g}\left| a\right|\sin\Phi+ \frac{\epsilon}{N}\left|  {J}^{-}\right| \right)\partial_{{r}}\left(\partial_\psi-\partial_\delta\right)\bar{P}
+\frac{2Nw+\left(N+2 {J}^{z}\right)\gamma_\phi -4 g \left| a\right| \left|  {J}^{-}\right|\cos\Phi }{8\left|  {J}^{-}\right|^{2}} \left(\partial_\psi-\partial_\delta\right)^{2} \bar{P} .
\label{eq:FP3}
\end{align}
Adiabatically eliminating the spin variables ${{r}},\mu$ and the phase difference $\delta$~\cite{carmichael2013statistical}, we derive an effective Fokker–Planck equation
\begin{equation}
    \partial_t\tilde{P}\left(z,\phi,t\right)\approx\left[\kappa_a\partial_zz+\frac{1}{2}D_a\partial^2_z+\frac{1}{2}D_\phi\partial^2_\phi+\frac{\chi\kappa_a}{\left| a\right|}\left(2 z-\frac{1}{2}\partial_z\right)\partial_\phi\right]\tilde{P}\left(z,\phi,t\right),
\label{eq:FPEff4}
\end{equation}
where $\tilde{P}\left(z,\phi,t\right)$ is the quasi-probability distribution for photons in amplitude–phase coordinates.
With the initial condition $\delta(z-z_0)\delta(\phi-\phi_0)$ at $t=0$ (which corresponds to a photon coherent state), this equation can be solved by the Green's function
\begin{equation}
\label{eq:Green}
    G(z,\phi,t;z_0,\phi_0,0)=\frac{1}{\sqrt{\pi\tilde{\sigma}^2}} \sum_{n=-\infty}^{\infty}\frac{e^{in(\phi-\phi_0)}}{2\pi}\exp\left[ -\frac{\tilde{D}_\phi}{2} n^2 t -\frac{\left(z-\tilde{z}_{n}\right)^2}{\tilde{\sigma}^2} +\tilde{c}_{n}\right],
\end{equation}
where
\begin{subequations}
\begin{align}
        \tilde{\sigma}^2 & \equiv \frac{D_a}{\kappa_a}\left(1-e^{-2\kappa_at}\right),
       \\
        \tilde{D}_\phi & \equiv {D_\phi}+\frac{4\chi^2D_a}{\left| a\right|^2}+\frac{2\chi^2\kappa_a}{\left| a\right|^2},
        \\
        \tilde{z}_{n} & \equiv z_0e^{-\kappa_a t}+i\frac{\chi D_a}{\left| a\right|\kappa_a}\left(1-e^{-\kappa_a t}\right)^2n+i\frac{\chi}{2\left| a\right|}\left(1-e^{-\kappa_a t}\right)n,
        \\ 
        \tilde{c}_{n} & \equiv -\frac{4i z_0 n \chi }{\left| a\right|}+\left(1-e^{-\kappa_at}\right)\left[\frac{D_a}{\kappa_a\left| a\right|^2}\left(3-e^{-\kappa_a t}\right)n^2\chi^2+i\frac{2z_0}{\left| a\right|}n\chi+\frac{1}{\left| a\right|^2}n^2\chi^2\right].
\end{align}
\end{subequations}
To visualize the photon distribution, we employ the $Q$-representation, defined as
\begin{align}
    Q(q,q*)&\equiv \frac{1}{\pi}\int_0^{2\pi} d\phi \int_{-\left| a\right|}^\infty dz\ e^{-\vert \alpha-q \vert^2}\tilde{P}(z,\phi).
    \label{eq:Q0}
\end{align}
The Q-representation of the Green's function solution in Eq.~\eqref{eq:Green} is
\begin{align}
   Q_{z_0,\phi_0} \left(q,q^*\right) \approx\frac{e^{-\left| q\right|^2}}{\pi {\sqrt{1+\tilde{\sigma}^{2}}}}\sum_{n=-\infty}^{\infty}e^{i n\left(\phi_q-\phi_0\right)-\frac{\tilde{D}_{\phi}}{2}n^{2}t+\tilde{c}_{n}-\frac{\left(\left| a\right|+\tilde{z}_{n}\right)^{2}}{1+\tilde{\sigma}^{2}}} \int_{0}^{2\pi}d\phi\frac{e^{in \phi}}{2\pi}
  e^{\frac{2\left(\left| a\right|+\tilde{z}_{n}\right)\vert q\vert\cos\phi+\vert q\vert^{2}\tilde{\sigma}^{2}\cos^{2}\phi}{1+\tilde{\sigma}^{2}}},\label{eq:Q1}
\end{align}
where $\phi_q\equiv \arg(q)$.
Here we have extended the integration domain in Eq.~(\ref{eq:Q0}) from $\left(-\left| a\right|,\infty\right)$ to $\left(-\infty,\infty\right)$ to facilitates normalization. 
The exponential factor scaling with $n^2$ accounts for the phase diffusion overtime. The term involving $\tilde{z}_{n}\ \vert q\vert\cos(\phi-\phi_{q})$ induces the phase–amplitude correlation, which manifest as asymmetric $Q$-distribution about the $\phi=0$ axis, giving rise to the twisting effect.
The $Q$-representation obtained from Eq.~(\ref{eq:Q1}) is shown in main text, which exhibit phase diffusion and twisting in the presence of spin–spin interactions. The twisting effect is a signature of quantum squeezing. 

\begin{figure}[htbp]
\includegraphics[scale=0.4]{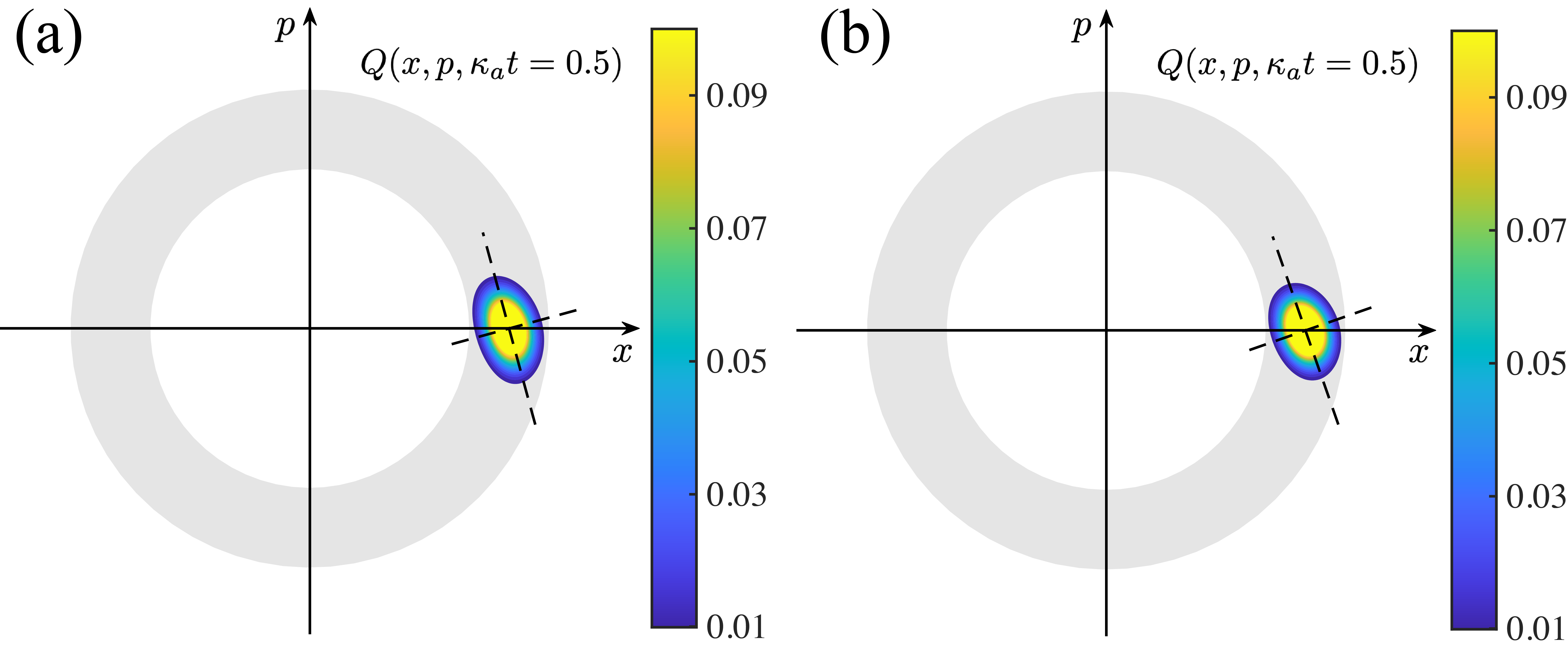}
\caption{\label{fig:figureS2}
Photon Q-distribution. (a) Calculated by solving the effective Fokker–Planck equation Eq.~(\ref{eq:FPEff4}). (b) Calculated by diagonalizing the Liouvilllian of the effective quantum model in Eq.~(\ref{eq:effMasterEq}) of main text. Parameters are the same as in Fig.~3a,b of main text. 
}
\end{figure}

\section{Effective master equation and Langevian equation}
To quantify the squeezing, we evaluate the quadrature fluctuation spectra. We first construct an effective master equation [Eq.~(\ref{eq:effMasterEq}) in main text] that reproduce the Fokker-Planck equation for a transient photon state initially in a coherent state, and then we derive the equivalent Langevian equation from the effective master equations. The quadrature fluctuation spectra are obtained from the Fourier transform of the Langevian equation.

It is straightforward to show that the effective master equation in Eq.~(\ref{eq:effMasterEq}) of main text leads to a Fokker–Planck equation the same as Eq.~\eqref{eq:FPEff4} for the short-time dynamics ($t\ll D_\phi^{-1}$) from the initial state $\tilde{P}\left(\alpha,\alpha^*,0\right)=\delta\left(\alpha-|a|\right)\delta\left(\alpha^*-|a|\right)$ (a coherent state with real photon amplitude). The $P$-representation of the model Eq.~(\ref{eq:effMasterEq}) of main text lead to the Fokker-Planck equation
\begin{align}
    \partial_t P\left(\alpha,\alpha^{*},t\right)=&\partial_{\alpha}\left\{\kappa_a\left(\alpha-\left| a\right|\right)+\frac{-i\chi\kappa_a}{2\left| a\right|^{2}}\left[\left(2\left| a\right|^{2}-1\right)-2\left|\alpha\right|^{2}\right]\alpha\right\}P+{\rm c.c.}
    \nonumber \\&+\frac{-i\chi\kappa_a}{2\left| a\right|^{2}}\left[\partial^2_\alpha\alpha^{2}-\partial^2_{\alpha^*}\left(\alpha^{*}\right)^{2}\right]P+\frac{1}{2}D_a\left(\partial^2_\alpha+\partial^2_{\alpha^*}+2\partial_\alpha\partial_{\alpha^*}\right)P
    \nonumber \\
    &+\frac{1}{2}D_\phi\left[\partial^2_\alpha\alpha^{2}-\partial_{\alpha}\alpha+\partial^2_{\alpha^*}\left(\alpha^{*}\right)^2-\partial_{\alpha^*}\alpha^{*}-2\partial_\alpha\partial_{\alpha^*}\left(\alpha\alpha^{*}\right)\right]P.
\end{align}
Linearizing around the mean-field solution $\alpha=\vert a\vert$ and keeping terms up to linear in $z$ and $\phi$, we obtain an effective Fokker-Planck equation analogous to Eq.~\eqref{eq:FPEff4}, namely,
\begin{equation}
    \partial_t\tilde{P}\left(z,\phi,t\right)\approx\left[\kappa_a\partial_zz+\frac{1}{2}D_a\partial^2_z+\frac{1}{2}D_\phi\partial^2_\phi+\frac{\kappa_a\chi}{\left| a\right|}\left(2 z-\frac{1}{2}\partial_z\right)\partial_\phi\right]\tilde{P}\left(z,\phi,t\right).\label{eq:FPEq2}
\end{equation}
Figure ~\ref{fig:figureS2} show that Eq.~\eqref{eq:FPEff4} and the effective master equation [Eq.~(\ref{eq:effMasterEq}) of main text] produce nearly the same $Q$-distribution.

The Langevian equation for the photon field can be derived from the effective master equation. We neglect the diffusion term ${\mathcal L}_{\hat{a}^\dagger\hat{a}}$ for $t\ll D_\phi^{-1}$ in the short-time dynamics. The Langevian equation for $\delta\hat{x}\equiv \left(\delta\hat{a}^{\dagger}+\delta\hat{a}\right)/\sqrt{2}$ and $\delta \hat{p}\equiv i\left(\delta\hat{a}^{\dagger}-\delta\hat{a}\right)/\sqrt{2}$ with $\delta\hat{a}\equiv \hat{a}-a$ {are}
\begin{subequations}
\begin{align}
    \frac{d\delta\hat{x}}{dt} & = -\kappa_a\delta\hat{x}+\sqrt{\kappa_a}\left[\hat{{f}}_1^\dagger(t)+\hat{{f}}_1(t)\right]+i2\sqrt{D_a}\left[\hat{{f}}_2^\dagger(t)-\hat{{f}}_2(t)\right],\\
     \frac{d\delta\hat{p}}{dt} & = -\kappa_a\delta\hat{p}+i\sqrt{\kappa_a}\left[\hat{{f}}_1^\dagger(t)-\hat{{f}}_1(t)\right]-2\chi \kappa_a\delta\hat{x}.
\end{align}
\end{subequations}
where $\hat{{f}}_i(t)$ (for $i=1,2$) are white-noise operators satisfying $\left\langle\hat{f}_i(t)\hat{f}_j(t')\right\rangle=0$, $\left\langle\hat{f}_i^\dagger(t)\hat{f}_j(t')\right\rangle=0$ and $\left\langle\hat{f}_i(t)\hat{f}_j^\dagger(t')\right\rangle=\delta_{ij}\delta(t-t')$. 
We can see that the interaction (the term associated with $\chi$) causes a `position'-dependent drift of the `momentum', inducing a twisting effect in the phase space.
Note that the Langevian equations above apply for the particular initial condition $\tilde{P}\left(\alpha,\alpha^*,0\right)=\delta\left(\alpha-|a|\right)\delta\left(\alpha^*-|a|\right)$, which is a coherent state in the $x$-axis. 
The fluctuation $\hat{f}_2$ corresponds to a dissipation of the position $\propto -D_a p$, which is zero for the particular initial condition. Consequently, it only contributes to the amplitude diffusion along the $x$-axis for the particular initial condition.

The symmetric correlation function is defined as
\begin{align} S_{AB}(\omega) & =\frac{1}{2}\int_{-\infty}^{\infty}d\tau e^{i\omega\tau}\left\langle \left\{ \delta\hat{A}(\tau),\delta\hat{B}(0)\right\} \right\rangle
\nonumber \\ & =\frac{1}{2}\left(\left\langle\delta\hat{A}\left(\omega\right)\delta \hat{B}\left(-\omega\right)\right\rangle +\left\langle \delta\hat{B}\left(-\omega\right)\delta\hat{A}\left(\omega\right)\right\rangle \right).
\nonumber
\end{align}
Solving the Langevian equations in the frequency domain by the Fourier transform, we obtain
\begin{align}
    &\left(\begin{array}{cc}
\left\langle \delta\hat{x}\left(\omega\right)\delta\hat{x}\left(-\omega\right)\right\rangle  & \left\langle \delta\hat{x}\left(\omega\right)\delta\hat{p}\left(-\omega\right)\right\rangle \\
\left\langle \delta\hat{p}\left(\omega\right)\delta\hat{x}\left(-\omega\right)\right\rangle  & \left\langle \delta\hat{p}\left(\omega\right)\delta\hat{p}\left(-\omega\right)\right\rangle 
\end{array}\right)\notag\nonumber\\
=&
\left(\begin{array}{cc}
-i\omega+\kappa_{a} & 0\\
2\chi\kappa_{a} & -i\omega+\kappa_{a}
\end{array}\right)^{-1}\left(\begin{array}{cc}
\kappa_{a}+4D_{a} & i\kappa_{a}\\
-i\kappa_{a} & \kappa_{a}
\end{array}\right)\left(\begin{array}{cc}
i\omega+\kappa_{a} & 2\chi\kappa_{a}\\
0 & i\omega+\kappa_{a}
\end{array}\right)^{-1},
\end{align}
which gives
\begin{equation}
   \Re\left[\left(\begin{array}{cc}
S_{xx}(\omega) & S_{xp}(\omega)\\
S_{px}(\omega) & S_{pp}(\omega)
\end{array}\right)\right]=\frac{4D_{a}+\kappa_{a}}{\kappa_{a}^{2}+\omega^{2}}\left(\begin{array}{cc}
1 & \frac{-2\kappa_{a}^{2}\chi}{\kappa_{a}^{2}+\omega^{2}}\\
\frac{-2\kappa_{a}^{2}\chi}{\kappa_{a}^{2}+\omega^{2}} & \frac{4\kappa_{a}^{2}\chi^{2}}{\kappa_{a}^{2}+\omega^{2}}+\frac{\kappa_{a}}{4D_{a}+\kappa_{a}}
\end{array}\right).\label{eq:NoiseM}
\end{equation}
By diagonalizing the real symmetric noise-spectrum matrix in Eq.~(\ref{eq:NoiseM}), we obtain two eigenvalues that define two principal spectra as
\begin{equation}
    S_{\pm}\left(\omega\right)=\frac{2\kappa_{a}}{\kappa_{a}^{2}+\omega^{2}}\left[\frac{2D_{a}+\kappa_{a}}{2\kappa_{a}}+\frac{4D_{a}\kappa_{a}\chi^{2}+\kappa_{a}^{2}\chi^{2}}{\kappa_{a}^{2}+\omega^{2}}\pm\sqrt{\left(\frac{4D_{a}\kappa_{a}\chi+\kappa_{a}^{2}\chi}{\kappa_{a}^{2}+\omega^{2}}\right)^{2}+\left(\frac{D_{a}}{\kappa_{a}}-\frac{4D_{a}\kappa_{a}\chi^{2}+\kappa_{a}^{2}\chi^{2}}{\kappa_{a}^{2}+\omega^{2}}\right)^{2}}\right].
\end{equation}
The amplitude diffusion can be neglected since $D_a\ll\kappa_a$. Therefore, we simplify the noise spectra as
\begin{equation}
    S_{\pm}\left(\omega\right)=\frac{\kappa_{a}}{\kappa_{a}^{2}+\omega^{2}}\left[1+\frac{2\kappa_{a}^{2}\chi}{\kappa_{a}^{2}+\omega^{2}}\left(\chi\pm\sqrt{1+\chi^{2}}\right)\right].
\end{equation}
The two principal spectra correspond to linear combinations of the fluctuation operators along the squeezed and anti-squeezed axes, i.e.,
\begin{subequations}
\begin{align}
    \hat{Q}_- & \equiv  \cos\left({\theta}/{2}\right)\delta\hat{x}+\sin\left({\theta}/{2}\right)\delta\hat{p},\\
    \hat{Q}_+ & \equiv -\sin\left({\theta}/{2}\right)\delta\hat{x}+\cos\left({\theta}/{2}\right)\delta\hat{p},
\end{align}
\end{subequations}
where
$$\theta=\tan ^{-1}\left({\chi}\right).$$ 
In the non-interacting limit ($\epsilon=0$, i.e., $\chi=0$), the two principal spectra are equal,
\begin{equation}
    \left. S_{+}(\omega)\right|_{\epsilon=0}=\left. S_{-}(\omega)\right|_{\epsilon=0}=\frac{\kappa_a}{\kappa_a^2+\omega^{2}}.
\end{equation}
In general, for interacting spins ($\epsilon\neq0$, i.e., $\chi\neq0$), the two principal spectra become unequal, i.e., $S_{-}\left(\omega\right)\neq S_{+}\left(\omega\right)$, and $S_{-}\left(\omega\right)$ can be smaller than $\left. S_{-}(0)\right|_{\epsilon=0}$. This indicates that the cavity field is squeezed by the spin–spin interaction. The decibel squeeze parameter at $\omega=0$ (which characterizing the stanationary squeezing) is
\begin{equation}
    \zeta(0)\equiv-10 \log_{10} \frac{S_-(0)}{\left. S_{-}(0)\right|_{\epsilon=0}}=-10\lg\left(1-2\vert\chi\vert\sqrt{1+\chi^2}+2\chi^2\right)=\frac{20}{\ln10}\operatorname{arcsinh}(\vert\chi\vert).
\end{equation}

\section{Exact Numerical Simulation}
For the $N$-spin master equation discussed in main text, the density matrix $\hat{\rho}$ has dimension ${\mathcal D}_{\hat{\rho}}=4^{N}\left(N_{{\rm cut}}+1\right)^{2}$, where $N_{{\rm cut}}$ is the cutoff of the cavity-photon Fock space. 
When superradiant lasing occurs, the mean intracavity photon number can reach the order of $N^{2}$.
Therefore, to ensure convergence of $\hat{\rho}$, the cutoff $N_{\rm cut}$ must be chosen at least of order $N_{\rm cut} \sim O\left(N^2\right)$.
Since ${\mathcal D}_{\hat{\rho}}$ grows exponentially with $N$, direct simulation of the master equation becomes infeasible for a large number of spins. We perform the exact numerical simulation to check the mean-field theory for a small $N$.

Noticing that the system is symmetric under the permutation of the spins, we expand the density matrix in the collective spin eigenstates and the photon number states as
\begin{equation}
\hat{\rho}=\sum_{n,n'=0}^{N_{{\rm cut}}}\sum_{\iota=1}^{D_J}\sum_{\iota'=1}^{D_{J'}} \sum_{J,M;J',M'}\rho_{J,M,\iota;J',M',\iota'}^{n,n'}\left\vert J,M,\iota\right\rangle\left\langle J',M',\iota'\right\vert\otimes\left\vert n\right\rangle\left\langle n'\right\vert,
\end{equation}
where $\left| n\right\rangle$ is the cavity state with $n$ photons, $\left\vert J,M,\iota\right\rangle$ are the spin eigenstates satisfying $\hat{J}^2\left\vert J,M,\iota\right\rangle=J(J+1)\left\vert J,M,\iota\right\rangle$ for $J\in\{0,1,\ldots,N/2\}$ and $\hat{J}^z\left\vert J,M,\iota\right\rangle=M\left\vert J,M,\iota\right\rangle$ for $M\in\{J,J-1,\ldots,-J$ with 
$\iota$ labelling the degenerate (multiplicity) subspaces, and $D_J$ denotes the multiplicity of the Dicke manifold $\left\vert J,M\right\rangle$, given by
\begin{equation}
D_J=\frac{N!\left(2J+1\right)}{\left({N}/{2}+J+1\right)!\left({N}/{2}-J\right)!}.
\end{equation}
If the initial state of the spins has no coherence between the degenerate subspaces labeled by different $\iota$ such as a fully polarized spin state, the pump process cannot generate cohrence between different subspaces. Thus, the density matrix at any time can be written as
\begin{equation}
\hat{\rho}=\sum_{n,n'}\sum_{J,M,M'}\rho_{J,M,M'}^{n,n'}\hat{P}_{J,M,M'}\otimes\left\vert n\right\rangle\left\langle n'\right\vert,\label{eq:rhoDicke}
\end{equation}
where
\begin{equation}
\hat{P}_{J,M,M'}=\frac{1}{D_J}\sum_{\iota=1}^{D_J}\left\vert J,M,\iota\right\rangle\left\langle J,M',\iota\right\vert.
\end{equation}
In the Dicke basis, the master equation of $\hat{\rho}$ becomes a set of linear equations of the coefficients $\rho_{J,M,M'}^{n,n'}$.
The effective dimension of $\hat{\rho}$, denoted as $\tilde{{\mathcal D}}_{\hat{\rho}}$, is determined by the total number of the coefficients $\rho_{J,M,M'}^{n,n'}$.
Considering that the cutoff of the cavity-photon Fock space scales as $N_{{\rm cut}} \sim O\left(N^{2}\right)$ and that $\rho_{J,M,M'}^{n,n'}=0$ if $M+n \ne M'+n'$ due to the $U(1)$ symmetry of the system, the effective dimension of $\hat{\rho}$  scales as $\tilde{{\mathcal D}}_{\hat{\rho}} \sim O\left(N^6\right)$. For feasible numerical simulation, we choose $N = 15$.

Now we derive the linear equations of $\rho_{J,M,M'}^{n,n'}$. 
The unitary evolution and the cavity loss conserve the total angular momentum $J$, so we can directly apply the Liouvillian super-operators on $\hat{P}_{J,M,M'}\otimes\left\vert n\right\rangle\left\langle n'\right\vert$ and obtain
\begin{align}
\left.\frac{d}{dt}\rho_{J,M,M'}^{n,n'}\right|_{H,\kappa}= & 
-i {\mathcal H}_{M,M'}^{n,n'}\rho_{J,M,M'}^{n,n'}
+
{\mathcal K}^{n+1,n'+1}\rho_{J,M,M'}^{n+1,n'+1} \nonumber \\
 & 
 +{\mathcal G}_{J,M-1}^{n+1}\rho_{J,M-1,M'}^{n+1,n'}
+{\mathcal G}_{J,M'-1}^{n'+1}\rho_{J,M,M'-1}^{n,n'+1}
-{\mathcal L}_{J,M+1}^{n-1}\rho_{J,M+1,M'}^{n-1,n'}
 -{\mathcal L}_{J,M'+1}^{n'-1}\rho_{J,M,M'+1}^{n,n'-1},
\end{align}
where $M,M'=-J,\cdots,J$, $J=0,1,2,\cdots,N/2$ for even $N$, $J=1/2,3/2,\cdots,N/2$ for odd $N$ and
\begin{subequations}  
\begin{align}
{\mathcal H}_{M,M'}^{n,n'}= & \omega_{z}\left(M-M'\right)-\frac{\epsilon}{N}\left[M^{2}-\left(M'\right)^{2}\right]+\omega_{\rm c}\left(n-n'\right)-i \frac{\kappa}{2}\left(n+n'\right),\\
{\mathcal K}^{n,n'}= & \kappa\sqrt{nn'},\\
{\mathcal G}_{J,M}^{n}= & {g}\sqrt{n}\sqrt{\left(J-M\right)\left(J+M+1\right)},\\
{\mathcal L}_{J,M,M'}^{n,n'}= & {g}\sqrt{n+1}\sqrt{\left(J+M\right)\left(J-M+1\right)},
\end{align}
\end{subequations}
where $\omega_z$ is the transition frequency of the spins.
The Liouvillian describing spin pump and relaxation does not conserve the total angular momentum. Using the relation of~\cite{chase2008collective}
\begin{align}
\sum_{j=1}^{N}\hat{s}_{j}^{\nu}\hat{P}_{J,M,M'}\left(\hat{s}_{j}^{\nu'}\right)^{\dagger}= & \frac{N+2}{4J\left(J+1\right)}A_{J,M}^{\nu}A_{J,M'}^{\nu'}\hat{P}_{J,M^{\nu},M'^{\nu'}}+\frac{N+2J+2}{4J\left(2J+1\right)}B_{J,M}^{\nu}B_{J,M'}^{\nu'}\hat{P}_{J-1,M^{\nu},M'^{\nu'}} \nonumber \\
 & +\frac{N-2J}{4\left(J+1\right)\left(2J+1\right)}C_{J,M}^{\nu}C_{J,M'}^{\nu'}\hat{P}_{J+1,M^{\nu},M'^{\nu'}},
\end{align}
with $\nu,\nu'\in\{+,-,z\}$ and 
\begin{subequations}
\begin{align}
 & M^{\pm}=M\pm1 \ {\rm and}\  M^{z}=M,\\
 & A_{J,M}^{\pm}=\sqrt{\left(J\mp M\right)\left(J\pm M+1\right)} \ {\rm and}\ A_{J,M}^{z}=M,\\
 & B_{J,M}^{\pm}=\pm\sqrt{\left(J\mp M\right)\left(J\mp M-1\right)}\ {\rm and} \ B_{J,M}^{z}=\sqrt{\left(J+M\right)\left(J-M\right)},\\
 & C_{J,M}^{\pm}=\mp\sqrt{\left(J\pm M+1\right)\left(J\pm M+2\right)} \ {\rm and} \ C_{J,M}^{z}=\sqrt{\left(J+M+1\right)\left(J-M+1\right)}.
\end{align}
\end{subequations}
we can obtain the evolution equation for $\rho_{J,M,M'}^{n,n'}$, that is,
\begin{align}
&\left.\frac{d}{dt}\rho_{J,M,M'}^{n,n'}\right|_{w,\gamma,\gamma_{\phi}} \nonumber\\
= & -{\mathcal X}_{J,M,M'}\rho_{J,M,M'}^{n,n'} + {\mathcal Y}_{J+1,M,M'}\rho_{J+1,M,M'}^{n,n'}+{\mathcal Z}_{J-1,M,M'}\rho_{J-1,M,M'}^{n,n'}
\nonumber \\
& +{\mathcal U}_{J,M+1,M'+1}\rho_{J,M+1,M'+1}^{n,n'}
+{\mathcal V}_{J+1,M+1,M'+1}\rho_{J+1,M+1,M'+1}^{n,n'}
+{\mathcal W}_{J-1,M+1,M'+1}\rho_{J-1,M+1,M'+1}^{n,n'}
\nonumber \\
 & +{\mathcal R}_{J,M-1,M'-1}\rho_{J,M-1,M'-1}^{n,n'}+{\mathcal S}_{J+1,M-1,M'-1}\rho_{J+1,M-1,M'-1}^{n,n'}+{\mathcal T}_{J-1,M-1,M'-1}\rho_{J-1,M-1,M'-1}^{n,n'}, 
\end{align}
where 
\begin{subequations}
\begin{align}
{\mathcal X}_{J,M,M'}= & \frac{w}{2}\left(N-M-M'\right)+\frac{\gamma}{2}\left(N+M+M'\right)+\frac{\gamma_\phi}{2}\left(N-\frac{MM'(N+2)}{J(J+1)}\right),\\
{\mathcal Y}_{J,M,M'}= & \frac{\gamma_\phi}{2}\frac{N+2J+2}{J(2J+1)}\sqrt{\left(J+M\right)\left(J-M\right)\left(J+M'\right)\left(J-M'\right)},\\
{\mathcal Z}_{J,M,M'}= & \frac{\gamma_\phi}{2}\frac{N-2J}{(J+1)(2J+1)}\sqrt{\left(J+M+1\right)\left(J-M+1\right)\left(J+M'+1\right)\left(J-M'+1\right)}
,
\\
{\mathcal U}_{J,M,M'}= & \frac{\gamma}{4}\frac{N+2}{J\left(J+1\right)}\sqrt{\left(J+M\right)\left(J-M+1\right)\left(J+M'\right)\left(J-M'+1\right)}\\
{\mathcal V}_{J,M,M'}= 
& \frac{\gamma}{4}\frac{N+2J+2}{J\left(2J+1\right)}\sqrt{\left(J+M\right)\left(J+M-1\right)\left(J+M'\right)\left(J+M'-1\right)},\\
{\mathcal W}_{J,M,M'}= & \frac{\gamma}{4}\frac{N-2J}{\left(J+1\right)\left(2J+1\right)}\sqrt{\left(J-M+1\right)\left(J-M+2\right)\left(J-M'+1\right)\left(J-M'+2\right)},\\
{\mathcal R}_{J,M,M'}= & \frac{w}{4}\frac{N+2}{J\left(J+1\right)}\sqrt{\left(J-M\right)\left(J+M+1\right)\left(J-M'\right)\left(J+M'+1\right)},\\
{\mathcal S}_{J,M,M'}= &\frac{w}{4}\frac{N+2J+2}{J\left(2J+1\right)}\sqrt{\left(J-M\right)\left(J-M-1\right)\left(J-M'\right)\left(J-M'-1\right)},\\
{\mathcal T}_{J,M,M'}= & \frac{w}{4}\frac{N-2J}{\left(J+1\right)\left(2J+1\right)}\sqrt{\left(J+M+1\right)\left(J+M+2\right)\left(J+M'+1\right)\left(J+M'+2\right)}.
\end{align}
\end{subequations}
Finally, the linear equations of $\rho_{J,M,M'}^{n,n'}$ are summarized as
\begin{align}
\frac{d}{dt}\rho_{J,M,M'}^{n,n'}= \left.\frac{d}{dt}\rho_{J,M,M'}^{n,n'}\right|_{H,\kappa}
+
\left.\frac{d}{dt}\rho_{J,M,M'}^{n,n'}\right|_{w,\gamma,\gamma_{\phi}}.
\end{align}
The steady state is obtained by imposing $d\rho_{J,M,M'}^{n,n'}/dt=0$, i.e., by finding the zero-eigenvalue eigenvector of the corresponding transformation matrix.

\begin{figure}[t]
\includegraphics[width=0.6\textwidth]{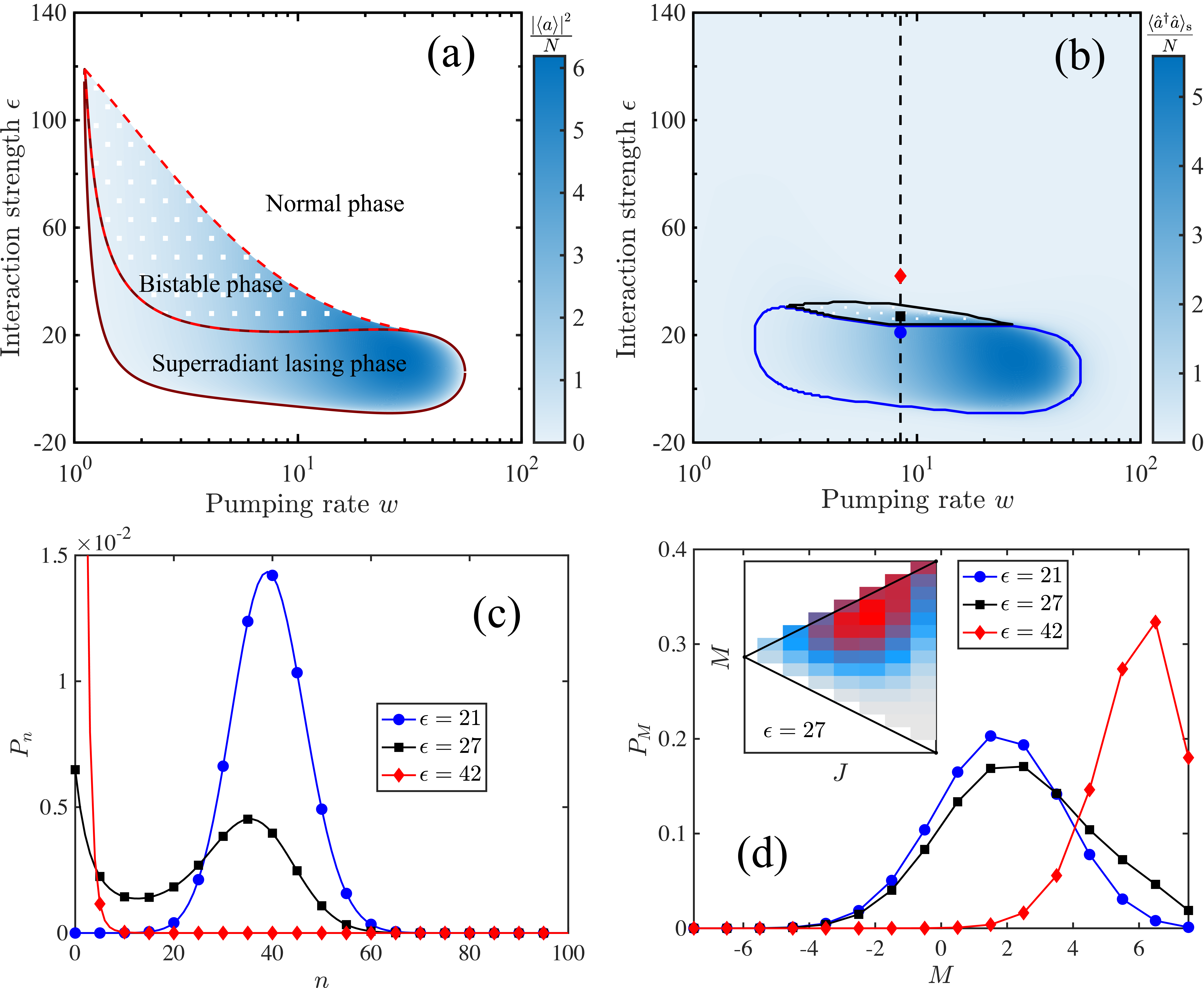}
\caption{\label{fig:figureS3} 
Comparison between exact numerical simulation and the mean-field theoretic results.
(a) Phase diagram given by the mean-field theory, as shown by the number of photons in the cavity. (b) Steady-state cavity photon number $N_{\rm ph} = \left\langle  a^\dagger a \right\rangle$ obtained by exact numerical simulation.  The number of spins is $N=15$. The lines are the phase boundaries roughly determined by the photon-number distribution. The three symbols mark the parameters for the three curves in (c) and (d) with the corresponding colors and symbols.
(c) Photon number distributions for various interaction strengths: an approximate exponential distribution for $\epsilon=42$, a Poisson distribution for $\epsilon=21$, and a mixture for $\epsilon=27$. 
(d) Distribution of spin polarization $P_M$ for three representative interaction strengths - approximately exponential for $\epsilon=42$, Poissonian for $\epsilon=21$, and a mixture for $\epsilon=27$. The inset shows the probability distribution in total spin and polarization $\left(J, M\right)$ for $\epsilon=27$. The parameters are: $\gamma=\gamma_\phi=\kappa=\Omega=1$ and $\Delta=6$.
}
\end{figure}

After obtaining the density matrix $\hat{\rho}$, we compute physical observables via $ \langle\hat{O}\rangle ={\rm Tr}\left[\hat{\rho}\hat{O}\right]$.
For an ensemble of $N$ spins, any collective-spin operator can be written as a linear combination of $\left(\hat{J}^{+}\right)^{p}\left(\hat{J}^{z}\right)^{r}\left(\hat{J}^{-}\right)^{q}$ with $p+r+q\leq N$ and $p,r,q$ being non-negative integers \cite{carmichael2013statistical}. The cavity photon operators can be expanded in the truncated Fock basis as linear combinations of $\left(\hat{a}^{\dagger}\right)^{k}\left(\hat{a}\right)^{k'}$ with $k,k'\leq N_{\rm cut}$.
Therefore, we have
\begin{equation}
\hat{O}=\sum_{k,k'=0}^{N_{{\rm cut}}}\sum_{p,r,q}^{p+r+q\leq N}C_{p,r,q}^{k,k'}\left(\hat{J}^{+}\right)^{p}\left(\hat{J}^{z}\right)^{r}\left(\hat{J}^{-}\right)^{q}\left(\hat{a}^{\dagger}\right)^{k}\left(\hat{a}\right)^{k'}.
\end{equation}
Then, all physical observables can be calculated by $\left\langle \hat{O}\right\rangle_{\rm s}=\sum_{k,k'=0}^{N_{{\rm cut}}}\sum_{p,r,q}^{p+r+q=N}C_{p,r,q}^{k,k'}B_{p,r,q}^{k,k'}$ with
\begin{align}
   B_{p,r,q}^{k,k'} 
\equiv & {\rm Tr}\left[\hat{\rho}_{\rm s}\left(\hat{J}^{+}\right)^{p}\left(\hat{J}^{z}\right)^{r}\left(\hat{J}^{-}\right)^{q}\left(\hat{a}^{\dagger}\right)^{k}\left(\hat{a}\right)^{k'}\right]\nonumber\\
= & \sum_{n,n'}\sum_{J,M,M'}\rho_{J,M,M'}^{n,n'}{\rm Tr}\left[\hat{P}_{J,M,M'}\vert n\right\rangle\left\langle n'\vert\left(\hat{J}^{+}\right)^{p}\left(\hat{J}^{z}\right)^{r}\left(\hat{J}^{-}\right)^{q}\left(\hat{a}^{\dagger}\right)^{k}\left(\hat{a}\right)^{k'}\right] \nonumber\\
= & \sum_{n,n'}\sum_{J,M,M'}\rho_{J,M,M'}^{n,n'}\left\langle J,M'\vert\left(\hat{J}^{+}\right)^{p}\left(\hat{J}^{z}\right)^{r}\left(\hat{J}^{-}\right)^{q}\vert J,M\right\rangle\left\langle n'\vert\left(\hat{a}^{\dagger}\right)^{k}\left(\hat{a}\right)^{k'}\vert n\right\rangle  \nonumber \\
= & \sum_{n,n'}\sum_{J,M,M'}\rho_{J,M,M'}^{n,n'}\left(M-q\right)^{r}\sqrt{\frac{n!}{\left(n-k'\right)!}\frac{n'!}{\left(n'-k\right)!}}\sqrt{\frac{\left(J+M\right)!}{\left(J-M\right)!}\frac{\left(J+q-M\right)!}{\left(J+M-q\right)!}}\nonumber\\
&~~~~~~~~~~~~\times\sqrt{\frac{\left(J+M'\right)!}{\left(J-M'\right)!}\frac{\left(J+p-M'\right)!}{\left(J+M'-p\right)!}}
\delta_{M'-p,M-q}\delta_{n'-k,n-k'}. 
\end{align}
Note that since the system has U(1) symmetry, any observable that does not satisfy the selection rule $p+k=q+k'$ vanishes in the steady 
state.

We perform exact numerical solution of the density matrix for a finite system of $N=15$ spins to validate the results obtained by the mean-field theory. 
As shown in Fig.~\ref{fig:figureS3}, the mean-field theoretic results are qualitatively reproduced by the exact numerical solution.
The sharp increase of the average cavity photon number $N_{\rm ph}=\left\langle \hat{a}^\dagger \hat{a}\right\rangle_{\rm s}$ (the dark blue region in Fig.~\ref{fig:figureS3}b) indicates the superradiant lasing. The photon number distribution (Fig.~\ref{fig:figureS3}c) and the spin polarization distribution (Fig.~\ref{fig:figureS3}d) are approximately exponential, Poissonian, or a mixture of the two, corresponding to the thermal, lasing, or bistable phase. The phase diagram extracted from the qualitative features of the photon-number distribution resemble those from the mean-field theory, as shown in Fig.~\ref{fig:figureS3}a, with discrepancies attributable to finite-size effects.

\bibliography{Article.bib}

\end{document}